\documentstyle[11pt,emulateapj]{article}
\tighten

\begin{document}

\def\sarc{$^{\prime\prime}\!\!.$}
\def\beginrefer{\section*{References}
\begin{quotation}\mbox{}\par}
\def\refer#1\par{{\setlength{\parindent}{-\leftmargin}\indent#1\par}}
\def\endrefer{\end{quotation}}

\title{ Caught in the Act; {\it Chandra} Observations of Microlensing of 
the Radio-Loud Quasar MG~J0414+0534}
\submitted{The Astrophysical Journal, accepted}

\author{G. Chartas\altaffilmark{1}, E. Agol\altaffilmark{2,3}, M. Eracleous\altaffilmark{1},
G. Garmire\altaffilmark{1}, M. W. Bautz\altaffilmark{4},
\& N. D. Morgan\altaffilmark{4}}


\altaffiltext{1}{Astronomy and Astrophysics Department, Pennsylvania State University,
University Park, PA 16802., chartas@astro.psu.edu}

\altaffiltext{2}{Theoretical Astrophysics, Caltech, Pasadena, CA 91125}

\altaffiltext{3}{{\it Chandra} Fellow}
\altaffiltext{4}{MIT Center for Space Research, 70 Vassar Street, Cambridge, MA 02139.}


\begin{abstract}
We present results from monitoring of the distant (z = 2.64),
gravitationally lensed quasar MG~J0414+0534 with the {\it Chandra X-ray
Observatory}.  An Fe~K$\alpha$ line at 6.49 $\pm$ 0.09~keV (rest-frame) with an
equivalent width of $\sim$ 190~eV consistent with fluorescence from a cold
medium is detected at the 99\% confidence level in the spectrum of the
brightest image A.  
During the last two observations of our monitoring
program we detected a five-fold increase of the equivalent width 
of a narrow Fe~K$\alpha$ line in the spectrum of image B but
not in the brighter image A whereas
image C is too faint to resolve the line.
The continuum emission component of image B
did not follow the sudden enhancement of the iron line
in the last two observations.
We propose that the sudden increase in the iron line strength from 
$\sim$ 190~eV to 900~eV only in image B can be
explained with a caustic crossing due to 
microlensing that selectively enhances a strip of
the line emission region of the accretion disk.
The non-enhancement of the continuum emission in the spectrum of image B
suggests that the continuum emission region is concentrated closer to the center of the
accretion disk than the iron line emission region
and the magnification caustic has not reached close enough
to the former region to amplify it.
A model of a caustic crossing event predicts discontinuities
in the light-curve of the magnification and provides
an upper limit of $\sim$ 5 $\times$ 10$^{-4}$ pc 
on the outer radius of the Fe~K$\alpha$ emission region.
The non-detection of any relativistic or Doppler shifts of the iron line 
in the spectrum of image B implies that the magnification 
caustic for the last two observations  
was located at a radius greater than $\sim$ 100 gravitational radii.

Each observation of the quadruply lensed quasar MG~J0414+0534 provides a view of the quasar at
four different epochs spaced by the time-delays between the lensed
images.  We produced a light-curve of the quasar X-ray flux by
normalizing the flux of each image to the mean flux of that image over
all observations.  We find significant deviations of the normalized
light-curve from unity especially for the faintest image C. A
plausible mechanism to explain the flux variability of image C is a
microlensing event.  Finally, spectral analysis of MG~J0414+0534
indicates the presence of significant absorption in excess of the Galactic value.
For absorption at the redshift of the lensed quasar we 
find an intrinsic column density of N$_{H}$ $\sim$ 5
$\times$ 10$^{22}$ cm$^{-2}$, consistent with the reddening observed
in the optical band.

\end{abstract}


\section{INTRODUCTION}
Direct imaging of the immediate environments of black holes in Active
Galactic Nuclei (AGNs) is beyond the capabilities of present day
telescopes; rough estimates of the characteristic sizes of emission
regions of AGN in the radio, optical and X-ray bands, based on
light-travel time arguments, correspond to angular sizes on the order
of tens of nano-arcseconds at a redshift of $z=1$. However, indirect
mapping methods have been developed recently to study the environs of
AGN without the need of nano-arcsec resolution.  One of these methods,
known as reverberation mapping (Blandford \& McKee 1982), has been
successfully used to determine the size and probe the structure of the
broad-line region in several AGN (see Peterson 1993 and Netzer \&
Peterson 1997 for recent reviews). This method relies on the time lag
between variations in the flux from a central source of ionizing
radiation and the response of the emission lines from photo-ionized gas
in the broad-line region. 
A closely related technique for indirect mapping of the accretion disk around
the black hole relies on the variation of the profiles of the Fe~K$\alpha$
fluorescence lines originating in the inner parts of the disk. The
line profiles are determined by a combination of the Doppler effect
and special and general relativistic effects (e.g., Fabian et al 1989;
Laor 1991) while their variations are caused by fluctuations in the
X-ray continuum that drive the line emission (Young \& Reynolds 2000).

A recently proposed, independent, method for indirect
imaging of AGN accretion disks exploits the high-magnification
microlensing events (HMEs) in gravitational lens (GL) quasars (Grieger
et al. 1988 and 1991; Schneider, Ehlers \& Falco 1992; Gould \& Gaudi
1997; Agol \& Krolik 1999; Yonehara et al. 1999; Mineshige \& Yonehara
1999, Yonehara 2001; Shalyapin 2001). During a microlensing event magnification
caustics produced by stars in the lensing galaxy traverse the plane of
the accretion disk and selectively magnify different emission
regions. An analysis of the light-curves of microlensing events
obtained in several wavelengths can be used to infer the surface
brightness and inclination angle of the accretion disk and possibly
provide constraints on the mass and spin of the black hole (e.g, Agol
\& Krolik 1999 and references within). A promising object for
application of this technique is the Einstein Cross.  With the first
detection of a microlensing event in the Einstein Cross by Irwin et
al. (1989) and subsequent verification by Corrigan et al. (1991) and
Ostensen et al. (1996) it became clear that future monitoring of the
Einstein Cross has the potential of yielding information on the
geometry of the accretion disk and possibly the spin and mass of the
black hole. Thus, a program to monitor magnification events
in the Einstein Cross has been undertaken as part of
the Optical Gravitational Lensing Experiment (OGLE, see Wo{\' z}niak et al. 1999 and
references within). Microlensing events have been detected with OGLE,
however a clear case of a HME due to a caustic crossing has yet to be
detected. Recently, constraints on the continuum source size of the lensed quasar
have been derived based on the microlensing light-curves of the Einstein
Cross (eg., Wyithe, Webster \& Turner 2000; Yonehara 2001; Shalyapin 2001).

In this paper we report the serendipitous detection of significant
variations in the shape and equivalent width (EW) of a reprocessed
Fe~K${\alpha}$ line only in one of the lensed images of the quasar
MG~J0414+0534. We interpret this change as the result of the crossing
of the accretion disk by a magnification caustic.

The gravitational lens system MG~J0414+0534 was discovered by Hewitt et
al. (1992). The system consists of four lensed images commonly referred
to as A1, A2, B and C (see Figure 1) with the separation between A1
and A2 being $\sim$ 0{\sarc}4.  The lensed source is a distant (z =
2.639), low-luminosity radio quasar (Lawrence et al. 1995).  A faint
blue arc is detected in HST observations (Falco et al. 1997) that
extends between images A1 and B. The arc is most likely the host
galaxy of the quasar lensed by the foreground elliptical galaxy
located at a redshift of 0.96 (Tonry \& Kochanek 1999).  An object
located $\sim$ 1~{\arcsec} West of image B and referred to as object X
(Schechter \& Moore 1993) possibly contributes to the lensing effect. IR observations of
MG~J0414+0532 indicate that the components of this system are
exceedingly red with respective R-H colors of 6.8 mag for image C,
3.2~mag for the arc, and 3.2 mag for the lens galaxy.  The origin of
the absorption is controversial with early studies mostly favoring the
dusty lens hypothesis (Lawrence et al. 1995; Malhotra, Rhoad, \&
Turner 1997; McLeod et al. 1998) and more recent work suggesting a
host galaxy origin (Tonry \& Kochanek 1999; Falco et al. 1997) or a
combination of lens and host (Angonin-Willaime et al. 1999).  The flux
ratios have been observed to be a function of wavelength and this has
been interpreted as either due to extinction in the lens or
microlensing of one or more of the images.  Recent VLBI and HST NICMOS
H-band images have been used to produce accurate lens models for this
system (Ros et al. 2000).  These models predict the time delays
between the lensed images to be ${\Delta}t_{AB}$ $\sim$ 16 $\pm$ 1.4
days and ${\Delta}t_{CB}$ $\sim$ 66 $\pm$ 5 days.  
The time-delays predicted from the lens model
of Ros et al. (2000) assume a flat cosmology with
$\Omega_{0}=1$ and $H_{0} = 65$ km s$^{-1}$ Mpc$^{-1}$.
The monitoring program of MG~J0414+0532 with {\it Chandra} was planned such that the
time interval between observations was substantially shorter than a
year, roughly the predicted average rise for a high magnification
event due to microlensing in this GL system (Witt, Mao \& Schechter 1995).  
A detailed description of the {\it Chandra} observations of
MG~J0414+0534 and the data reduction procedures used is presented in
section 2.
The resolved X-ray lensed images of this GL system
are presented in section 3.
The spectral analysis revealing the properties of 
an intervening absorber and reprocessed iron line features 
is detailed in section 4. We conclude with a discussion 
of the origin of the detected iron line features
and the prospects of mapping the accretion disk of a distant 
quasar with microlensing of a reprocessed Fe~K${\alpha}$ line.

\section{OBSERVATION AND DATA REDUCTION}

MG~J0414+0534 was observed with the Advanced CCD Imaging spectrometer
(ACIS) instrument (G. P. Garmire et al. 2002, in preparation) 
onboard the {\it Chandra X-ray
Observatory} (CXO) in a series of five pointings listed in Table
1. The pointing of the telescope placed MG~J0414+0534 on the
back-illuminated S3 chip of ACIS.

The observations were performed as part of a survey of GL systems with
the CXO aimed at finding suitable candidates for time-delay
measurements.  The software package CIAO 2 provided by the {\it
Chandra X-ray Center } was used to process the data. We chose to
remove a $\pm$ 0{\sarc}25 randomization applied to the event positions
in the CXC processing. The randomization is used in the standard CXC
pipeline processing to deal with aliasing effects noticed in
observations with durations of less than $\sim$ 4~ks.  Removal of the
pixel randomization results in an improvement of the spatial resolving
power by about 0{\sarc}1. To improve the spatial resolution 
of ACIS even further we employed a method
recently developed by Tsunemi et al. (2001).
The physical basis of this method is as follows;
Any photon that is detected by ACIS produces a charge cloud of electrons
that is collected by one or more pixels. The different arrangements of charge
are referred to as ACIS grade distributions.  Particular grade
distributions often referred to as corner events can
be used to locate the position of the incident photons to sub-pixel accuracy.
By correcting the positions of corner pixel events the spatial resolution
of ACIS is significantly improved.

\normalsize

\section{RELATIVE ASTROMETRY AND PHOTOMETRY}

The {\it Chandra} image of MG~J0414+0534 obtained from combining the five observations
listed in Table 1 is presented in Figure 1.
Images A1 and A2 are not clearly resolved in the {\it Chandra} image.
We applied the Lucy - Richardson (L-R) maximum likelihood
deconvolution technique to improve the image quality. L-R
deconvolution was applied to the combined image of the five
observations of MG~J0414+0534. For the deconvolution we supplied a
point spread function (PSF) created by the simulation tool \verb+MARX+ (Wise
et al. 1997).  The X-ray spectrum used to generate the PSF was that
determined from our spectral analysis of the combined images of
MG~J0414+0534. In particular, we used an absorbed power-law model with
absorption of N$_{H}$ = 10.85 $\times$ 10$^{20}$ cm$^{-2}$ due to our
Galaxy, intrinsic absorption at z = 2.64 of N$_{H}$ = 4.76 $\times$
10$^{22}$ cm$^{-2}$ , and a photon index of 1.72.  In Figure 1 we
present the original and deconvolved image of the combined
observations.  In Table 1 we provide the epochs, exposure times and
detected events for the lensed images for each observation of
MG~J0414+0534.  In Table 2 we present the relative X-ray and optical
(HST) positions of MG~J0414+0534 images A1, A2, and B with respect to
image C. We find a good agreement between optical and X-ray positions.

The detected count rates of images A1+A2, B, and C versus Julian date
are shown in Figure 2. Figure 2 also shows the X-ray flux 
ratios A/B, A1/A2 and C/B for each observation. 
These flux ratios are not phase-corrected for the predicted time-delays.
The flux ratios A/B and C/B were derived 
by extracting events from circles of radii 1~{\arcsec} centered on 
the images. Due to the close separation between images A1 and A2
of $\sim$ 0{\sarc}4 we estimated the X-ray flux ratio A1/A2 by 
performing a likelihood analysis on the binned event positions; 
We modeled A1 and A2 with simulated PSF's and 
minimized the Cash (1979) C statistic formed between the
observed and predicted events to find the 
best fit normalization values. The relative positions
of A1 and A2 were held fixed to the optical positions
listed in Table 2. For the 5th observation of MG~J0414+0534
with the highest signal-to-noise ratio  we obtain
a A1/A2 X-ray flux ratio of 1.3 $\pm$ 0.2 which
is consistent within the uncertainties to the radio flux ratio of 1.1. 
In Table 3 we present the 8GHz (Katz, Moore, Hewitt 1997), the
H, R, and I band (CfA-Arizona Space Telescope LEns Survey (CASTLES))
and the 2-8keV band flux ratios of the components of MG~J0414+0534.

For a comparison with other observations we have included in Figure 2 the 
HST values of the flux ratios obtained in the optical H-band.
In Figure 3 we present the {\it Chandra} light-curve of MG~J0414+0534.  
The count-rates of each image have been normalized to the average 
count-rate of that image for
the five observations listed in Table 1.  Based on the lens models of
Ros et al. (2000) we assume that image B leads images A and C by 16 and
66~days, respectively. Long term variability is significant for image
C and not in the other images suggesting
that microlensing may be present in image C. 
The double peaked light-curve of MG~J0414+0534
is suggestive of a double-caustic crossing event in image C.  


\section{SPECTRAL ANALYSIS}

We performed spectral analysis for each 
observation of MG~J0414+0534 listed in Table 1.
A variety of spectral models were fitted to the data 
employing the software tool \verb+XSPEC v11+ (Arnaud 1996).
The spectra of images A, B, and C were extracted from circular regions
centered on the images with radii of 1~{\arcsec}.
The background was determined by extracting events within an
annulus centered on the midpoint between images A and B and 
with inner and outer radii of 5~{\arcsec} and
30~{\arcsec}, respectively. All derived errors are at the 90\% 
confidence level unless quoted otherwise.		

We began with spectral models consisting of simple power-laws with
Galactic absorption due to neutral cold gas with a column density of
N$_{H}$ = 0.1085 $\times$ 10$^{22}$ cm$^{-2}$.  The models also
include extra-galactic absorption placed at either the redshift z =
0.96 of the lens or at the redshift z = 2.64 of the lensed quasar.
Our spectral fits require absorption in addition to
Galactic, however, these fits cannot constrain the redshift of the
absorbers. The results of these fits are shown in Table 4.  No significant
variability of the spectral slope and intervening column density
towards image A is detected with the exception of a possible change in
absorbing column towards image A between the first and second
observation.  In Figure 4 we show the 68\% and 90\% confidence levels
between the best fit spectral slope and intrinsic absorption for
observations I and II. The apparent change in $\Gamma$ and
intrinsic absorption between these observations is significant 
at the $\sim$ 68\% level.

To increase the signal-to-noise ratio we fitted the combined spectrum of all
lensed images A, B and C for the five observations. For intrinsic
absorption at a redshift of z = 2.64 we find best fit values for the
spectral slope and column density to be $\Gamma$ =
1.72$^{+0.08}_{-0.09}$ and N$_{H}$(z=2.64) = 4.76 $\times$ 10$^{22}$
cm$^{-2}$ respectively (fit 21 of Table 4).  The spectral slope
of MG~J0414+0534 is consistent with the mean index of $\sim$ 1.66 $\pm$
0.04 and dispersion of $\sigma$ = 0.22 found for radio-loud quasars
observed with ASCA (Reeves \& Turner 2000).  

The combined spectra of all images and of image A alone for all
observations are shown in Figure 5.  A strong line feature is detected
at $\sim$ 1.78~keV.  We investigated whether this feature is produced by the
known Si~K fluorescence line of the ACIS instrument and concluded that 
the feature is not an instrumental effect. We estimate the total 
number of background counts in the extraction
radius of 1~{\arcsec} used for image A to be 0.9 counts in the 0.3 to
10~keV band and only 0.08 counts are estimated to lie in the range of
1.5-2 keV. To illustrate the effect of background 
and the instrumental Si~K line we combined background spectra from all the
observations of MG~J0414+0534 with large extraction annuli centered on
MG~J0414+0534 with inner and outer radii of 5~{\arcsec} and 100~{\arcsec}.
In Figure 6 we show the combined spectrum of image A
with the combined background spectrum scaled with appropriate normalization
for the apertures used to extract the spectra.
We conclude that the instrumental contamination from the
Si~K fluorescence line to the {\it background subtracted} spectra of
the images of MG~J0414+0534 is negligible.

We added a Gaussian line component to the absorbed power-law model and
present the best fit parameters in Table 5.  We find a rest frame
energy of the line of 6.49 $\pm$ 0.09~keV consistent with
reprocessed Fe K$\alpha$ emission from neutral material.  In Figure 7
we show the confidence contours of the iron line energy versus line
normalization for the fit to the combined spectrum of all images over
all observed epochs.  The line is detected at a high confidence level
($ > $99\%).  The best fit value for the rest frame equivalent width
of the iron line in image A is 193$^{+123}_{-117}$ eV (90\% confidence levels) 
(fit 1 of Table 5) with a flux in the line of 4.4 $\times$ 10$^{-15}$ erg s$^{-1}$ cm$^{-2}$.
The estimated unlensed and
unabsorbed 2-10 keV X-ray luminosity of MG~J0414+0534 is
6.9 $\pm$ 0.4 $\times$ 10$^{44}$ erg s$^{-1}$ (H$_{0}$ = 50 km
s$^{-1}$ Mpc$^{-1}$, q$_{0}$ = 0.5).  
We have assumed a magnification factor of 30 based on the 
detailed modeling of this system by Trotter et al. (2000).

We searched for the presence of a Compton reflection ``hump'' by
adding the \verb+XSPEC+ model \verb+PEXRAV+ (Magdziarz \& Zdziarski 1995) 
that simulates Compton reflection of an exponentially cut
off power law from neutral material.  The strength of the reflection
component is parameterized with the reflection scaling factor $R =
\Omega/2\pi$, where $\Omega$ is the solid angle subtended by the
reflection material from the source. For upper cut-off energies of the
primary X-ray spectrum of E$_{c}$ = 100~keV and 500~keV we place 90\%
upper limits of the reflection scaling factor of R $ < $ 0.44 and R $
< $ 0.32, respectively.

During observations 4 and 5 we detect a significant increase in the intensity 
of the iron line in image B (see Figure 8).
The Fe~K$\alpha$ line is detected in the individual
spectra of image B for observations 4 and 5. Due to the low
S/N, however,  of the individual spectra the constraints
of the EW of the Fe~K$\alpha$ line are poor.
To increase the S/N of the spectrum of image
B observations 4 and 5 were combined. 
The residuals of a spectral fit to the combined spectrum 
are suggestive of the presence of a broader
component blueshifted with respect to the narrow line. Specifically, a spectral fit
using a model that consists of 2 Gaussian lines, a simple power-law
modified by Galactic and intrinsic absorption yields a significant
improvement at the $ > $ 99\% confidence level based on the F-test
compared to a fit that did not include the Gaussian lines
(see fits 1 and 3 of Table 6).  The 68.3\%, 90\% and 99\% confidence
levels between the line energy of the narrow component versus the
normalization of this component are shown in Figure 7.  The best fit
value for the narrow line energy is 6.48 $\pm$ 0.11~keV consistent
with emission of neutral iron of a cold medium with an equivalent
width of 906$^{+604}_{-520}$~eV (fit 3 of Table 6). The broad line is centered at an energy
of 9.21$^{+1.18}_{-0.77}$~keV with a width of 1.3 $\pm$ 0.5~keV. Quoted
energies, equivalent widths, and line widths are in the rest frame of
the quasar.  As we discuss in section 5, the combined
spectrum of image B for observations 1, 2 and 3 is
consistent with the presence of an Fe line with 
an equivalent width similar to that of image A.

\section{DISCUSSION AND CONCLUSIONS}

Assuming that the observed X-ray absorption is intrinsic to the quasar
we expect that a fraction of the observed Fe~K$\alpha$ emission is
produced by fluorescence of the intrinsic absorber.  Adopting a
covering fraction, $f_{c}$, of unity and assuming a small optical depth an
upper limit of the EW (rest frame) produced by the intervening absorber
is provided by the following expression (Halpern 1982):
\begin{equation}
{EW \sim 140f_{c}\left({{N_{H}}\over{1.6 \times 10^{23}}}\right)
\left({{A_{\rm Fe}}\over{4 \times 10^{-5}}}\right)~{\rm eV}},
\end{equation}
where, $A_{\rm Fe}$ is the iron abundance with respect to
hydrogen. For the best fit value of an intrinsic absorber (fit 21
Table 2) of $N_{\rm H} = 4.8 \pm 0.7 \times 10^{22}~{\rm
cm}^{-2}$ and assuming cosmic iron abundances we estimate that only 
$\sim$ 40~eV of the observed 190~eV
can be attributed to fluorescence from the intervening absorber.  We
conclude that most of the narrow iron Ka line observed in image A of
MG~J0414+0534 is produced by reprocessing from a cold medium.
Our estimate also assumes an ``average'' iron abundance relative to
Hydrogen of 4 $\times$ 10$^{-5}$. Values reported in the literature 
(Anders \& Grevesse, 1989; Feldman U. 1992; Anders \& Ebihara 1982) 
range from 3.2 $\times$ 10$^{-5}$ to 4.7 $\times$ 10$^{-5}$, therefore our adopted value is
in the middle of the range, which extends about 20\% above and below.

The EW of the Fe~K${\alpha}$ line is expected to scale linearly with the 
reflection scaling factor R and be a function of the inclination angle of
the accretion disk and the spectral slope $\Gamma$ of the incident
power-law spectrum. For the observed spectral slope of 
$\Gamma$ $\sim$ 1.6 (fit 3 Table 5)
the equivalent width approximately scales as 120R eV,
assuming an inclination angle of 60$^{\circ}$ (e.g., George \& Fabian 1991).
The spectral fits in Table 5 place an upper limit 
of R$\sim$ 0.5 (at the 90\% confidence limit) for an 
assumed inclination angle of 60$^{\circ}$.
Spectral fit 3 of Table 5 was repeated for inclination angles ranging
between 20$^{\circ}$ and 80$^{\circ}$. We find an upper limit on
$W_{K\alpha}(i,\Gamma)$R of $\sim$ 60~eV, where
$W_{K\alpha}(i,\Gamma)$ is the predicted
value of the equivalent width of the iron line for an inclination angle i,
and a spectral slope $\Gamma$ assuming cosmic abundances for Fe
(see figure 14 in George \& Fabian 1991).
The observed value of EW $\sim$ 190~eV of the Fe line in image A,
appears to be too large for the observed limit of R.
A plausible explanation for the large EW in image A is that
iron may be over-abundant in MG~J0414+0534. A high abundance of Fe would
strengthen the Fe line and at the same time make the Compton reflection hump weaker
by increasing the opacity above the Fe~K${\alpha}$ edge.
A high abundance of Fe will also result in an increased 
contribution of the fluorescence of the intrinsic absorber as described
in equation 1. However, it is not clear whether a high
Fe abundance in the cold reprocessing medium must be accompanied by
a high Fe abundance in the intrinsic absorber.
Alternatively, a high iron equivalent width may be created by
emission which is stronger in the direction of the reprocessing cloud
than in the direction of the observer, or by a variable source which
was stronger in the past.

The observed weakness of the Compton reflection component
in the radio-loud quasar MG~J0414+0534 by a factor of 2 
compared to values of R observed in (radio-quiet) Seyfert galaxies 
is in agreement with recent observations of radio-loud 
AGN's (Wo{\' z}niak et al. 1998; Sambruna, Eracleous, \& Mushotzky 1999;
Eracleous, Sambruna, \& Mushotzky 2000). In particular,
Eracleous et al. (2000) propose that the inner regions of radio-loud objects
may contain a quasi-spherical ion torus or an advection-dominated
accretion flow that would result in small
solid angles subtended by the disk to the primary X-ray source.

The properties of the iron line feature observed in image B appear to
be different than those observed in the brighter image, A.
Specifically, the line feature in image A appears to be present in all
observations and is considerably weaker than the
line detected in image B for the last two observations ($\sim$ 910~eV). 
In particular, for the combined spectra of the first 
three observations we find the equivalent widths of the iron lines
in the spectra of images A and B to be
273$_{-177}^{+178}$~eV and $222_{-222}^{+334}$eV, respectively.
For the combined spectra of the last two
observations we find the equivalent widths of the iron lines
in the spectra of images A and B to be
114$_{-114}^{+168}$~eV and 906$_{-520}^{+604}$~eV, respectively.
The iron line profile in image A does not show any broad component above 6.4~keV.


Given that the light from image B leads that of image A by about 16
days one would expect to have observed the narrow and broad iron line
components seen in image B also in image A during observation 
5. It is therefore unlikely that the iron line observed in image B is
produced by a temporal increase in the Fe~K$\alpha$ reprocessed
component, since such an event would have been observed in image A as
well. 

A plausible explanation of the detection of a strong iron line in
image B and not in image A is the occurrence of a microlensing event
in image B beginning sometime between the third and fourth
observation.  We estimate the size of the projected Einstein ring
radius, $\zeta_{\rm E} = [(4GM_{star}/c^{2})(D_{\rm os}D_{\rm ls}/D_{\rm
ol})]^{1/2}$, produced by a star of mass $M_{star}$ in the lens plane,
where D represents the angular diameter distances, and the subscripts
{\it l, s}, and {\it o} refer to the lens, source, and observer,
respectively.  For the GL system MG~J0414+0534 with lens and source
redshifts of $z_{\rm lens} = 0.96$ and $z_{\rm source} = 2.64$
respectively, and assuming an isolated star of mass {\it M} the
Einstein-ring radius on the source plane is $\zeta_{\rm E} \sim 0.01
(M_{star}/{\rm M}_{\odot})^{1/2}$~pc.  This is of the order of 
the line emitting region of an AGN with a black 
hole mass of $\sim 10^{8}~{\rm M}_{\odot}$.
For our microlensing simulations we assumed a straight fold caustic
traversing the accretion disk. This assumption implies 
that the sizes of the X-ray continuum emission and iron line reprocessing
regions are much smaller than the projected Einstein radius of the
perturbing star. X-ray variability studies of quasars indicate
that the size of the X-ray continuum emission region is of
the order of $\sim$ 1 $\times$ 10$^{-4}$ pc (eg., Chartas et al. 2001).
Also recent estimates (Wyithe et al. 2000; Yonehara 2001; Shalyapin, 2001) 
on the continuum source size in the
Einstein Cross based on the analysis of a microlensing event in this
system indicate that the size of the continuum emission region is less
than 1 $\times$ 10$^{15}$ cm. There are also numerous reports of similarly rapid
variability in Seyfert galaxies, which place similar upper limits on
the size of the X-ray source. So far, variability studies 
of iron lines in the spectra of radio
loud quasars are scarce and the geometry of the iron line
emission region in these objects is not well constrained.
However, the profiles of the Fe~K$\alpha$ lines of luminous Seyfert galaxies and
intermediate-luminosity quasars are quite broad and asymmetric (eg.,
Nandra et al. 1997), suggesting very strongly that the lines originate
in the inner parts of an accretion disk where the gravitational field
is quite strong.  Model fits to these line profiles indicate a
characteristic size of the line-emitting region of $10-1000~r_{\rm g}$, 
where, $r_{g} = {{GM}/{c^{2}}}$, is the gravitational radius.
For a 1 $\times$ 10$^{8}$ M$_{\odot}$ black hole this range of sizes
for the line-emitting region 
corresponds to $\sim$ 4.8 $\times$ 10$^{-5}$~pc - 4.8 $\times$ 10$^{-3}$~pc.
We conclude that the sizes of the X-ray continuum and
iron line regions in the radio-loud
quasar MG~J0414+0534 are likely to be smaller
than the estimated projected Einstein radius,
thus, making our assumption of a straight fold caustic
a reasonable one.
Microlensing events are produced by a star or a group of stars in the
foreground lensing galaxy. Such events will not lead to time-delayed
magnification in the remaining images and therefore could explain the
non-detection of a strong Fe~K$\alpha$ line profile in image A during
observation 5. A microlensing event could also explain the
large equivalent width Fe K$\alpha$ line (EW $\sim$ 910~eV) observed
in image B during observation 4 and 5.  As the caustic network
produced by the stars in the lensing galaxy traverses the accretion
disk, regions of the disk near the caustics will be magnified.  For a
caustic fold crossing one would expect selective magnification of a
strip of the disk as shown in Figure 9, and a corresponding change in
the spectrum.  The magnification of a point source close to a fold
caustic scales as the inverse square root of its distance to the
caustic.  Specifically, the amplification of a point source by a 
straight fold caustic is:

\begin{equation}
{A = A_{0} + {{K}\over{(x - x_{c})^{1/2}}}H(x - x_{c})},
\end{equation}

\noindent
where, $A_{0}$ is the amplification outside the caustic,
$K$ is the caustic amplification factor, $x$ is the coordinate
perpendicular to the caustic in units of $\zeta_E$, $x_{c}$ is the position of the caustic along 
the x-axis and $H(x - x_{c})$ is the Heaviside function. 
The caustic amplification factor is of the order of
$K/A_0 = \beta \left(\zeta_E\right)^{1/2}$, where $\beta$
is a constant of order unity
(see equation 7 in Witt, Kayser, \& Refsdal 1993).
The probability distribution for K has also been computed by
Wyithe \& Turner (2001), who show that at optical depth of order unity the
microlensing magnification patterns produced
by a mass function of microlenses can be approximately
reproduced by considering a mean microlens mass. 
We have computed 
the distribution of $K/A_0$ for $\kappa=0.4635$ and $\gamma=0.3253$,
appropriate for image B in the model by Ros et al. (2000), using the
code of Wyithe \& Turner (2001).  We find that if the entire optical depth
is due to stars, then the average value of $\beta$ is $\langle \beta\rangle=
0.2\pm 0.18$.  We note that there is a factor of 2 discrepancy between the 
predictions of $\langle K\rangle$ by Witt (1990) and Wyithe \& Turner (2001).
Given the large uncertainties, we will scale our results with $\beta$.

A plausible model that can explain the observed amplification
of only the iron line component and not the continuum X-ray emission
places the caustic at the time of the observed amplification
within the iron line emission region. 
To explain the non-amplification of the continuum emission 
during observations 4 and 5 we postulate that the 
thermal emission region of the disk and the Compton up-scattered emission 
region of the hard X-ray source lie within smaller radii than the 
iron line reprocessing region.


To assess the plausibility of the above hypothesis, we estimate the
magnification due to the caustic crossing needed to produce the factor
of $\sim$ 4.7 $\pm$ 2.3 (1 $\sigma$ errors) 
increase in the equivalent width of the Fe~K$\alpha$
line. For the purpose of carrying out this rough estimate, we adopt
the simple disk geometry shown in Figure~9.  We assume that the
line is emitted within an annulus region of the accretion disk,
with a radius of $r_{\rm out}$ and the continuum source is confined to be within $r_{\rm in}$
and it can be either a sphere (e.g., an ADAF) or a point source at or
above the center of the disk (e.g., as assumed for Seyfert
galaxies). The important point here is that we take the continuum
source to be smaller than the emission line source.

Figure 10 shows the amplification of the iron line region by 
a straight fold caustic as a function of time.
For this simple case we have assumed a flat emissivity profile.
As the caustic crosses one side of the ring the brightness
jumps by a factor of $f = I_{f}/I_{in} \sim (A_0 + K/(2r)^{1/2})/A_0$,
where, $I_{in}$ and $I_{f}$ are the iron line fluxes from the ring
before and after the caustic crosses the first edge, and
$r=(r_{out}+r_{in})/2$ is the center of the annulus.
The reason for this jump is that flux integrated in the y direction (parallel
to the caustic) has a cusp at the edges of the ring which goes to
infinity if the ring is infinitely thin.  As the caustic crosses this
edge, a discontinuity is created.  For a ring with a finite
thickness, this discontinuity becomes a smooth rise with a width
equal to $2dr/v_c$, where $v_{c}$ is the caustic velocity
and $dr = (r_{out}-r_{in})/2$.
In Figure 10 we show the amplification of the ring emission for 
$dr/r$ ratios of 0.025, 0.1, 0.5 and 1 (the last case is a uniform
disk of emission). After the discontinuity, 
more of the ring becomes amplified,
rising to a peak as the caustic crosses the other edge of the ring. 
In practice, the maximum amplification is given by:

\begin{equation}
{A_{max} = A_{0} + {{K}\over{{\pi}(2r)^{1/2}}} { \left[ 1 + \ln\left({{64r}\over{dr}}\right) \right]  }}
\end{equation}

If one assumes the change in the flux of the 
iron line is due to a caustic crossing the
disk, then the first jump condition, assuming $dr \ll r$ 
places an upper bound on the radius $r$ of the ring of, 

\begin{equation}
r < {\beta^2\zeta_E \over 2 (f-1)^2} 
\end{equation}

\noindent
on the outer radius of the emission region. 
For a typical value of $M_{star} = 1 M_{\odot}$
and for the observed amplification of the iron line
of a factor of 4.7$\pm$ 2.3 we find $r_{out} <  10^{-4}(\beta/0.2)^2$pc 
at the 68$\%$ confidence level.

If the observed increase in the flux of the iron line
in image B is occurring during the peak of the amplification 
curve then the upper bound on the ring radius is,

\begin{equation}
{r} <  {{{ {\beta^2{\zeta_{E}}\over{2(f-1)^{2}{\pi^{2}}}}{\left[ 1 + \ln\left({{{64r}\over{dr}}}\right)      \right]^{2} } }}} 
\end{equation}

The constraint on the radius of the ring for the case of maximum
amplification depends on $\ln({{r}/{dr}})$ and is therefore
only weakly dependent on the width of the ring. For
$dr/r$ = 0.1 the second case leads to an upper bound of 
the ring radius of $r < 6 \times 10^{-4}(\beta/0.2)^2 $pc at the 68$\%$ confidence level.
The large equivalent width of the iron line implies a large
covering factor, which may require $dr \sim r$.  In this case, the
maximum magnification is $A_{max}=A_0+ K/r^{1/2}$ 
(Witt et al. 1993), yielding an
upper limit which is twice as large as the thin annulus jump condition. 
With the limited sampling of the amplification curve we cannot 
definitively distinguish which of the two cases corresponds to the 
observed increase in the 
iron line flux of image B. However, the case of maximum amplification
is expected to last for a relatively short interval ($\sim$ days)
whereas the 4th and 5th observations 
of MG~J0414+0534 are separated by $\sim$ 80 days.
We therefore suggest that the observed amplification of the Fe~K${\alpha}$ line is
most likely caused by the initial crossing of the 
ring-shaped emission region by the caustic.
The non-detection of any amplification of the continuum emission 
in image B for observations 4 and 5 implies that the caustic 
during these observations had not crossed the second side
of the ring. Had the caustic already crossed the center of the disk it 
would have magnified any emission, including the continuum emission, behind the caustic.
This asymmetry in microlensing by a caustic is described by the
Heaviside function in equation 2. 
The amplification of the iron line changed
by a factor of $\sim$ 5 over a period of about 30 days,
giving a derivative of $\sim$ 40 mag per year.  This derivative is
large enough to imply that the observed change 
in the EW of the Fe~K${\alpha}$ image B
is likely due to a high-amplification event, meaning that the
Fe~K${\alpha}$ region is amplified by a caustic or cusp.
A limit on the source size can also be obtained
by calculating the probability of such a large change in amplification
in image B over a short timescale as a function of source size.
Such a calculation will be presented in a future publication.


Based on our simulations of caustic crossings 
we also expect to detect significant distortions of the
iron line due to special and general relativistic effects and Doppler effects
if the iron line emission region extends below $\sim$ 100 $r_{g}$.
``Snapshot'' spectra from detailed simulations for the
case where the iron line emission region extends below $\sim$ 100 $r_{g}$
are shown in Figure 11. The simulation assumes a flat, face-on disk about
a Schwarzchild black hole with turbulent velocity $\sigma_v$ equal to one
percent of the Keplerian velocity.
The caustic strength used for the simulations shown 
in Figure 11 is $K/A_0 = 10r_{g}^{1/2}$.
These simulations assume an $r^{-3}$ emissivity profile and include relativistic and Doppler effects. 
Based on these iron line simulations and the observed 
width and energy of the iron line we place a
lower bound of 100 $r_{g}$ on the distance of the 
caustic from the center of the disk.
Assuming the fold caustic at 100 $r_{g}$ we 
estimate that the strength of the caustic 
needed to explain the observed 4.7$\pm$2.3 increase in the 
EW of the iron line is 
$K/A_0 \sim \beta \left({\zeta_{E}}\right)^{1/2} = (27\pm 13)r_{g}^{1/2}$. 
This estimate of the caustic strength was derived assuming that
the un-microlensed equivalent width of the broad line is equal to 190~eV, and
that the disk covers 2${\pi}$ of the continuum source.
This relation provides an estimate of the mass of the black hole
as a function of the average mass of the lensing stars.
Specifically, if we assume the average mass of the stars responsible for the microlensing
event to lie between 0.1 M$_{\odot}$  and 1 M$_{\odot}$ we estimate the mass
of the black hole in MG~J0414+0534 to range between 
$3.6\times 10^{6}(\beta/0.2)^2$ M$_{\odot}$ and 
$1.1\times 10^{7}(\beta/0.2)^2$ M$_{\odot}$, 
respectively.  If the caustic is actually at a larger
radius than $100 r_g$, then a larger amplification is required, implying
a smaller mass black hole. Remarkably, this indicates that the X-ray 
luminosity is near Eddington if the emission is unbeamed and isotropic 
and if the average microlens mass is near solar.

Recently, there has been considerable interest in using
the lensing effect to probe for dark matter substructure
in galaxy halos (Metcalf \& Madau 2001; Chiba 2001).
To address the question of whether such substructure can be
responsible for the observed microlensing event in image B
we provide an estimate of the mass of the lensing object.
Assuming the quasar is emitting isotropically with an observed
unlensed luminosity lying between
1 and 0.1 times the Eddington luminosity
we estimate the mass of the black hole to lie between
5.3 $\times$ 10$^{6}$ and 5.3 $\times$ 10$^{7}$  M$_{\odot}$,
respectively. Using the constraint,
$\zeta_{E}^{1/2} = (27\pm13)r_{g}^{1/2}/\beta$, derived
from our analysis of the microlensing event in MG~J0414+0534
we find that the effective mass of the perturber
lies in the range of $0.2(\beta/0.2)^{-4}$ M$_{\odot}$ and 
$20(\beta/0.2)^{-4}$ M$_{\odot}$. 
We conclude that the lensing object is most likely
a star or group of stars and not a dark satellite in the lens galaxy.
A second argument that rules against substructures as the cause
of the microlensing event is that the
timescale for substructure to change
the X-ray fluxes is much longer than the
microlensing timescale by individual stars.

If the iron line in MG~J0414+0534 is concentrated
in a ring with $r_{inner} > 100r_{g}$
we predict to see no GR distortion as the caustic traverses the accretion disk and
only a change in the intensity of the line is expected.
This picture in which the iron line region is 
confined to large radii is supported by the results of X-ray 
observations of nearby radio-loud AGNs (see
Wo\'zniak et al. 1998 and Eracleous, Sambruna, \& Mushotzky 2000, and
references therein). Microlensing offers an independent
way of testing this scenario.

The timescale for the microlensing event in image B is of the order 
of $t_{e} = R_{s}/v$, where, $R_{s}$ is the size of the lensed source region
and $v$ is the velocity of the caustics in the lens plane (measured in our time
frame) given by equation (B9) in Kayser, R., Refsdal, S., \& Stabell, R. (1986). 
If we assume an observer velocity of 360~km/s (this is measured 
from the CMB dipole), lens and AGN velocities of 600~km/s,
and sum the three-dimensional velocities in equation (B9) in RMS, 
we find a typical velocity of a caustic of $\sim$
170~km/s and a typical timescale of the microlensing event of
$t_{e}$ $\sim 2 (R_s/10^{15}{\rm cm})$ years.

An apparent puzzle from our present observations is the predicted frequency of
a caustic crossing in MG~J0414+0535 of order
$\zeta_{E}/v$ $\sim$ 60 ($v$/170~km/s)$^{-1}$ $(M_{star}/M_{\odot})^{1/2}$ years.
There are several uncertain factors in this frequency estimate of order
unity, the main ones being the spacing of caustics perpendicular and
parallel to the shear and the optical depth of the lens.
However, there are four images in MG~0414+0535 and there are about ten
lenses that have been observed in the X-ray band
with {\it Chandra}, so perhaps overall the
detection of a microlensing event in image B is not so surprising.

If we continue to monitor MG~J0414+0534 in the following years
we predict that the profile of the iron line will significantly change
if the iron emission region of the radio loud quasar MG~J0414+0534
extends below 100$r_{g}$. If the iron line in MG~J0414+0534 is concentrated
in a ring with $r_{inner} > 100r_{g}$
we expect to see no GR distortion and the EW of the line in image B
should return to the un-microlensed value of $\sim$ 190eV. 

We also predict that in the upcoming months the caustic
will approach the center of the accretion disk
and magnify the continuum emission region.
The anticipated duration of magnification of the continuum region
depends on several factors that include
the present distance of the caustic from the continuum emission region,
the velocity of the caustic and the size of continuum emission region.
Several deeper {\it Chandra} observations of MG~J0414+0534 planned
for the following year will allow us to possibly observe
the onset of the magnification of the continuum emission region
and provide a constraint on its size.

We would like to thank Pat Broos for
providing the TARA software package, Koji Mori for providing  
subpixel correction software and the anonymous referee 
for providing many useful comments and suggestions. 
We thank Stuart Wyithe for use of his microlensing code and
Jose Mu\~noz for providing lensing model parameters. 
We acknowledge financial support by NASA grant NAS 8-38252.
Support for EA was provided by the National Aeronautics and Space Administration through
{\it Chandra} Postdoctoral Fellowship Award Number PF0-10013 issued by the
{\it Chandra} X-ray Observatory Center, which is operated by the Smithsonian
Astrophysical Observatory for and on behalf of the National Aeronautics
Space Administration under contract NAS8-39073.

\newpage

\scriptsize
\begin{center}
\begin{tabular}{lllllllc}
\multicolumn{8}{c}{TABLE 1} \\
\multicolumn{8}{c}{{\it Chandra} Observations of MG~J0414+0534} \\
& & & & & & &  \\ \hline\hline
\multicolumn{1}{l} {Observation} &
\multicolumn{1}{c} {Obsid} &
\multicolumn{1}{c} {Exposure} &
\multicolumn{1}{c} {Roll Angle} &
\multicolumn{1}{c} {R$_{A1+A2}$$^{a}$} &
\multicolumn{1}{c} {R$_{B}$$^{b}$} &
\multicolumn{1}{c} {R$_{C}$$^{c}$} &
\multicolumn{1}{c} {R$_{Bkg}$$^{d}$} \\
Date               &     & Time &         &      &      &      & \\
                   &     &      &         &      &      &      & \\
                   &     &  s   & (${}^{\circ}$)  &10$^{-3}$ cnts s$^{-1}$  & 10$^{-3}$  cnts s$^{-1}$ & 10$^{-3}$ cnts s$^{-1}$& 10$^{-6}$ cnts $('')$$^{-2}$ s$^{-1}$   \\ \hline
2000-01-13& 417 & 6578 &293.248& 51.4 $\pm$ 2.8 & 13.5 $\pm$ 1.4 & 3.0 $\pm$ 0.7  & 4.6 $\pm$ 0.5  \\
2000-04-02& 418 & 7437 &267.869& 48.3 $\pm$ 2.6 & 12.9 $\pm$ 1.3 & 7.7 $\pm$ 1.0  &27.2 $\pm$ 1.2  \\
2000-08-16& 421 & 7251 &102.926& 49.0 $\pm$ 2.6 & 14.3 $\pm$ 1.4 & 6.1 $\pm$ 0.9  & 3.4 $\pm$ 0.4\\
2000-11-16& 422 & 7504 & 40.134& 55.6 $\pm$ 2.7 & 12.9 $\pm$ 1.3 & 7.9 $\pm$ 1.0  & 4.0 $\pm$ 0.5 \\
2001-02-05&1628 & 9020 &277.900& 55.4 $\pm$ 2.5 & 16.8 $\pm$ 1.4 & 5.9 $\pm$ 0.8  & 5.0 $\pm$ 0.5  \\
\hline \hline
\end{tabular}
\end{center}
NOTES:\\
{}$^{a}$ $R_{A1+A2}$ are the detected event rates from images A1 and A2 of MG~J0414+0534
extracted from circular regions centered on the mid-point between A1 and A2
with radii of 1$''$. Only events with standard ASCA grades 0,2,3,4,6
and energies lying between 0.2 and 10keV were selected. \\

\noindent
{}$^{b}$ $R_{B}$ are the detected event rates from image B of MG~J0414+0534
extracted from circular regions centered on B
with radii of 1$''$.\\

\noindent
{}$^{c}$ $R_{C}$ are the detected event rates from image C of MG~J0414+0534
extracted from circular regions centered on C
with radii of 1$''$.\\

\noindent
{}$^{d}$ $R_{Bkg}$ are the detected background event rates per arcsec$^{2}$ extracted
from annuli centered on MG~J0414+0534 with inner and outer radii of 5$''$ and 30$''$ respectively.
Only events with standard ASCA grades 0,2,3,4,6 were extracted. \\

\clearpage
\scriptsize
\begin{center}
\begin{tabular}{lllll}
\multicolumn{5}{c}{TABLE 2}\\
\multicolumn{5}{c}{OPTICAL AND X-RAY OFFSETS OF MG~J0414+0534 IMAGES} \\
& & & & \\ \hline\hline
\multicolumn{1}{c} {Telescope} &
\multicolumn{1}{c} {B  } &
\multicolumn{1}{c} {A1} &
\multicolumn{1}{c} {A2} &
\multicolumn{1}{c} {C}  \\
\multicolumn{1}{c} {} &
\multicolumn{1}{c} {$\Delta\alpha$($''$), $\Delta\delta$($''$)} &
\multicolumn{1}{c} {$\Delta\alpha$($''$), $\Delta\delta$($''$)} &
\multicolumn{1}{c} {$\Delta\alpha$($''$), $\Delta\delta$($''$)} &
\multicolumn{1}{c} {$\Delta\alpha$($''$), $\Delta\delta$($''$)} \\ \hline
HST     &0,0   &0.60$\pm$0.01,-1.94$\pm$0.01 &0.73$\pm$0.01,-1.55$\pm$0.01 &1.345$\pm$0.01, 1.64$\pm$0.01    \\
{\it Chandra} &0,0   &0.57$\pm$0.02, -1.92$\pm$0.02 & 0.69$\pm$0.02,  -1.53$\pm$0.02    &1.30$\pm$0.02, 1.63$\pm$0.02    \\
\hline \hline
\end{tabular}
\end{center}

NOTE-\\
${}^{a}$ Offsets in RA and Dec with respect to image C. Relative positions taken with the
{\it Hubble Space Telescope} (HST) are from Angonin-Willaime et al. (1999). \\

\clearpage
\small
\begin{center}
\begin{tabular}{llll}
\multicolumn{4}{c}{TABLE 3}\\
\multicolumn{4}{c}{ MULTI-WAVELENGTH FLUX RATIOS }\\
\multicolumn{4}{c}{OF MG~J0414+0534 COMPONENTS}\\
 & & &\\ \hline\hline
\multicolumn{1}{c} {Waveband} &
\multicolumn{1}{c} {A1/A2} &
\multicolumn{1}{c} {A/B} &
\multicolumn{1}{c} {C/B}\\ \hline
8GHz             & 1.114 $\pm$ 0.002 & 4.882 $\pm$ 0.004 & 0.384 $\pm$ 0.001\\
H band           & 1.36 $\pm$ 0.05 & 4.45 $\pm$ 0.15 & 0.46 $\pm$ 0.02\\
R band           & 4.00 $\pm$ 0.06 & 2.65 $\pm$ 0.08 & 0.52 $\pm$ 0.04  \\
I band           & 2.64 $\pm$ 0.06 & 3.06 $\pm$ 0.07 & 0.48 $\pm$ 0.04\\
2-8keV           & 1.3 $\pm$ 0.2 & 3.75 $\pm$ 0.20 & 0.47 $\pm$ 0.04\\ \hline \hline
\end{tabular}
\end{center}

NOTE-\\
8GHz flux ratios are taken from the VLA data of Katz, Moore \& Hewitt (1997).
The H, R, and I band data are taken from 
the CfA-Arizona Space Telescope LEns Survey (CASTLES) of gravitational lenses
website {\it http://cfa-www.harvard.edu/glensdata/}.
Error bars for the X-ray data are at the 68\% confidence level.

\clearpage

\scriptsize
\begin{center}
\begin{tabular}{llllllcl}
\multicolumn{8}{c}{TABLE 4} \\
\multicolumn{8}{c}{RESULTS FROM SPECTRAL FITS OF ABSORBED POWER-LAW MODELS TO THE } \\
\multicolumn{8}{c}{INDIVIDUAL SPECTRA OF IMAGES OF MG~J0414+0534 } \\
& & & & & & &\\ \hline\hline
\multicolumn{1}{c} {Fit$^{a}$} &
\multicolumn{1}{c} {Epoch} &
\multicolumn{1}{c} {Image} &
\multicolumn{1}{c} {$\Gamma$} &
\multicolumn{1}{c} {$N_{H}(z=0.96)$} &
\multicolumn{1}{c} {$N_{H}(z=2.64)$} &
\multicolumn{1}{c} {Flux$^{b}$} &
\multicolumn{1}{c} {$\chi^{2}_{\nu}(\nu)$} \\
 &             &          &            & $10^{22}$ cm$^{-2}$ & $10^{22}$ cm$^{-2}$  & $10^{-13}$ erg s$^{-1}$ cm$^{-2}$ &     \\ \hline
 &             &          &            &                    &                     & &  \\
1&I &A + B + C &1.89$_{-0.25}^{+0.28}$ &    ...                & 4.50$_{-1.60}^{+2.17}$ &4.0 & 0.7(17)  \\
2&I &A + B + C &1.94$_{-0.26}^{+0.30}$ &1.02$_{-0.36}^{+0.50}$ & ...  & 3.9& 0.68(17)  \\
3&I &A         &2.12$_{-0.30}^{+0.34}$ &    ...             & 6.74$_{-2.13}^{+2.76}$ & 2.5 & 0.47(11)  \\
4&I &A         &2.22$_{-0.33}^{+0.38}$ &1.57$_{-0.50}^{+0.66}$ & ...             & 2.4 & 0.36(11) \\
 &             &          &            &                    &                     & &  \\ \hline
 &             &          &                                 &                     &                     & &  \\
5&II &A + B + C  &1.58$_{-0.20}^{+0.22}$  & ...             & 3.17$_{-1.15}^{+1.48}$& 5.4 & 0.99(21) \\
6&II &A + B + C  &1.61$_{-0.21}^{+0.24}$  & 0.71$_{-0.26}^{+0.34}$ & ...    & 5.3 & 0.97(21) \\
7&II &A          &1.43$_{-0.30}^{+0.33}$  & ...          & 2.25$_{-1.42}^{+1.94}$  &4.2 & 0.75(12) \\
8&II &A          &1.45$_{-0.31}^{+0.44}$        &0.50$_{-0.32}^{+0.36}$         & ...          &4.1 & 0.76(12) \\
 &             &           &                               &                    &                     & &  \\ \hline
 &             &          &                                &                    &                     & &  \\
9&III &A + B + C  &1.68$_{-0.27}^{+0.30}$  & ...     & 4.09$_{-1.51}^{+1.94}$  & 5.3 & 0.75(20)  \\
10&III &A + B + C  &1.73$_{-0.27}^{+0.31}$  & 0.90$_{-0.32}^{+0.42}$   & ...   & 5.2 & 0.75(20) \\
11&III &A          &1.66$_{-0.38}^{+0.42}$  &  ...        &3.92$_{-2.02}^{+2.81}$   & 3.4 & 1.08(11) \\
12&III &A          &1.71$_{-0.40}^{+0.47}$  & 0.87$_{-0.45}^{+0.66}$  & ...    & 3.2& 1.08(11) \\
 &             &           &                                &                    &                     & &  \\ \hline
 &             &          &                                &                    &                     & &  \\
13&IV &A + B + C  &1.66$_{-0.21}^{+0.22}$   &  ...      & 4.50$_{-1.39}^{+1.77}$  & 5.9 & 0.89(24)  \\
14&IV &A + B + C  &1.71$_{-0.23}^{+0.25}$   & 1.02$_{-0.33}^{+0.43}$  & ...      & 5.7 & 0.85(24) \\
15&IV &A          &1.92$_{-0.29}^{+0.33}$     &  ...    &6.14$_{-2.12}^{+2.59}$  & 3.2 & 0.86(14) \\
16&IV &A          &2.00$_{-0.34}^{+0.40}$     & 1.42$_{-0.52}^{+0.73}$  & ... & 3.2 & 0.83(14) \\
 &             &           &            &                                        &                     & &  \\ \hline
 &             &          &            &                                        &                     & &  \\
17&V &A + B + C   & 1.74$_{-0.16}^{+0.17}$         & ...         & 4.98$_{-1.17}^{+1.45}$   & 5.9 & 0.80(31) \\
18&V &A + B + C   & 1.78$_{-0.16}^{+0.18}$         & 1.11$_{-0.26}^{+0.33}$        & ...          & 5.8& 0.76(31) \\
19&V &A   & 1.66$_{-0.22}^{+0.25}$       & ...         & 5.14$_{-1.58}^{+2.01}$ & 4.3 & 0.85(19)  \\
20&V &A   & 1.70$_{-0.24}^{+0.27}$       & 1.14$_{-0.35}^{+0.45}$   &  ... & 4.2& 0.85(19) \\
 &             &           &                                &                    &                     & &  \\ \hline
 &             &          &                                &                    &                     & &  \\
21&ALL &A + B + C  & 1.72$_{-0.09}^{+0.08}$       & ...    & 4.76$_{-0.74}^{+0.70}$ & & 1.22(108)  \\
22&ALL &A + B + C  & 1.76$_{-0.09}^{+0.09}$       & 1.07$_{-0.17}^{+0.16}$ & ... & & 1.21(108)  \\
23&ALL &A   & 1.71$_{-0.11}^{+0.11}$       & ...         & 4.77$_{-0.92}^{+0.92}$ & & 0.91(74)  \\
24&ALL &A   & 1.76$_{-0.12}^{+0.11}$       &  1.08$_{-0.21}^{+0.20}$         & ... & & 0.89(74)  \\
25&ALL &B   & 1.70$_{-0.28}^{+0.31}$       & ...         & 3.25$_{-2.33}^{+2.56}$ & & 0.74(17)  \\
26&ALL &B   & 1.76$_{-0.17}^{+0.33}$       & 0.79$_{-0.54}^{+0.59}$         & ...  & & 0.74(17)  \\
27&ALL &C   & 1.63$_{-0.26}^{+0.1.6}$       & ...         & 3.9$_{-2.1}^{+2.9}$ & & 0.40(6)  \\
28&ALL &C   & 1.65$_{-0.28}^{+0.34}$       & 0.87$_{-0.47}^{+0.67}$  & ... & & 0.35(6)   \\
 &             &          &                             &                    &                     & &  \\
\hline \hline
\end{tabular}
\end{center}
\noindent
NOTES-\\
All models include Galactic absorption due to neutral cold gas with a column density of
N$_{H}$ = 0.1085 $\times$ 10$^{22}$ cm$^{-2}$. All derived errors are at the 90\% 
confidence level. The reduced chi-squared is defined as $\chi^{2}_{\nu}$ = $\chi^{2}/{\nu}$, where ${\nu}$,
the number of degrees of freedom, is given in parentheses.\\
$^{a}$ The spectral fits were performed within the energy ranges 0.4 - 6~keV \\
$^{b}$ Flux is estimated in the 2-10keV band \\

\clearpage

\scriptsize
\begin{center}
\begin{tabular}{llllllllll}
\multicolumn{10}{c}{TABLE 5} \\
\multicolumn{10}{c}{RESULTS OF FITS THAT INCORPORATE } \\
\multicolumn{10}{c}{COMPTON REFLECTION AND/OR IRON LINE EMISSION MODELS} \\
& & & & & & & & & \\ \hline\hline
\multicolumn{1}{c} {Fit} &
\multicolumn{1}{c} {Model} &
\multicolumn{1}{c} {$\Gamma$} &
\multicolumn{2}{c} {Reflection} &
\multicolumn{4}{c} {Iron Line} &
\multicolumn{1}{c} {$\chi^{2}_{\nu}(\nu)$} \\
 &&&&&&&&&  \\
\multicolumn{1}{c} {} &
\multicolumn{1}{c} {} &
\multicolumn{1}{c} {} &
\multicolumn{1}{c} {R} &
\multicolumn{1}{c} {E$_{c}$} &
\multicolumn{1}{c} {E$_{line}$} &
\multicolumn{1}{c} {EW} &
\multicolumn{1}{c} {FWHM} &
\multicolumn{1}{c} {F$_{line}$} &
\multicolumn{1}{c} {} \\
 &                      &                      &             & (keV)     &  (keV)                 & (eV)           &(km s$^{-1}$)& (erg s$^{-1}$ cm$^{-2}$)&      \\  \\ \hline
 &                      &                      &             &           &                        &                &             &                        &          \\
1&pl + line             &1.70$_{-0.10}^{+0.10}$&             &           & 6.49$_{-0.09}^{+0.09}$ &193$_{-117}^{+123}$&$ < $ 12000   &  4.4 $\times$ 10$^{-15}$   & 0.89(75)       \\
2&pl + line             &1.72$_{-0.09}^{+0.09}$&             &           & 6.47$_{-0.08}^{+0.07}$ &218$_{-94}^{+101}$&$ < $ 12000   &  7.6 $\times$ 10$^{-15}$   & 1.22(109)       \\
3&pl + line + reflection&1.62$_{-0.06}^{+0.09}$& $ < 0.44 $  &  100$^{f}$& 6.4$^{f}$              &191$_{-93}^{+93}$&             & 6.7 $\times$ 10$^{-15}$   & 1.22 (110)     \\
4&pl + line + reflection&1.67$_{-0.05}^{+0.05}$& $ < 0.32 $  &  500$^{f}$& 6.4$^{f}$              &198$_{-92}^{+99}$&             & 6.9 $\times$ 10$^{-15}$   & 1.22 (110) \\
& &   &  &          &                           &                      &                          &     &       \\
\hline \hline
\end{tabular}
\end{center}
\noindent
NOTE-\\
Fit 1 is performed to the combined spectra of image A for all epochs.
Fits 2, 3, and 4 are performed to the combined spectra of all images for all epochs.
$E_{line}, EW,$ and $FWHM$ are calculated in the quasar rest frame.
The symbol $f$ denotes that the quantity is fixed during the spectral fitting process.
All quoted errors are at the 90\% confidence level.
Spectral models for fits contain Galactic absorption
fixed at N$_{H}$ = 0.1085 $\times$ 10$^{22}$ cm$^{-2}$
and intrinsic absorption set as a free parameter.

\clearpage

\scriptsize
\begin{center}
\begin{tabular}{llllllllll}
\multicolumn{10}{c}{TABLE 6} \\
\multicolumn{10}{c}{RESULTS OF FITS TO IMAGE B THAT INCORPORATE } \\
\multicolumn{10}{c}{IRON LINE EMISSION MODELS} \\
& & & & & & & & & \\ \hline\hline
\multicolumn{1}{c} {Fit} &
\multicolumn{1}{c} {Model} &
\multicolumn{1}{c} {$\Gamma$} &
\multicolumn{3}{c} {Narrow Line} &
\multicolumn{3}{c} {Broad/Disk Line} &
\multicolumn{1}{c} {$\chi^{2}_{\nu}(\nu)$} \\
 &&&&&&&&&  \\
\multicolumn{1}{c} {} &
\multicolumn{1}{c} {} &
\multicolumn{1}{c} {} &
\multicolumn{1}{c} {E$_{line}$} &
\multicolumn{1}{c} {EW} &
\multicolumn{1}{c} {$\sigma$} &
\multicolumn{1}{c} {E$_{line}$} &
\multicolumn{1}{c} {EW} &
\multicolumn{1}{c} {$\sigma$ or Inc} &
\multicolumn{1}{c} {} \\
 &                       &                       & (keV)                 & (eV)                &  (keV)                  & (keV)                 & (keV)                &                    &      \\  \\ \hline
 &                       &                       &                       &                     &                        &                        &                      &                          &          \\
1&pl                     &1.70$_{-0.14}^{+0.15}$ & ...                   &  ...                & ...                    & ...                    & ...                  & ...                      & 1.41(26)       \\ 
2&pl + 1 gline           &1.69$_{-0.29}^{+0.36}$ & 6.48$_{-0.10}^{+0.10}$&880$_{-516}^{528}$  & $ < $ 0.15             & ...                    &...                   & ...                      & 1.25(23)       \\ 
3&pl + 2 glines          &1.99$_{-0.39}^{+0.49}$ & 6.48$_{-0.11}^{+0.11}$&906$_{-520}^{+604}$  & $ < $ 0.15             &9.21$_{-0.77}^{+1.18}$  &5.50$_{-2.2}^{+6.8}$  & 1.3$_{-0.5}^{+0.5}$keV    & 0.94(20)       \\ 
4&pl + diskline          &1.81$_{-0.36}^{+0.45}$ & 6.46$_{-0.09}^{+0.13}$&786$_{-464}^{+565}$  & $ < $ 0.15             &8.84$_{-0.59}^{+1.85}$  &3.4$_{-1.3}^{+5.4}$  & 45$^{\circ}$ $^{f}$          & 1.02(21)       \\ 
& &   &  &          &                            &                       &                          &     &       \\
\hline \hline
\end{tabular}
\end{center}
\noindent
NOTE-\\
Fits 1,2,3 and 4 are performed on the combined spectrum
of image B for observations 4 and 5.
All quoted errors are at the 90\% confidence level.
$E_{line}, EW,$ and $\sigma$ are calculated in the quasar rest frame. 
Spectral models for fits contain Galactic absorption
fixed at N$_{H}$ = 0.1085 $\times$ 10$^{22}$ cm$^{-2}$
and intrinsic absorption set as a free parameter.
The symbol $f$ denotes that the quantity is fixed during the spectral fitting process.


\clearpage

\normalsize

\beginrefer

\refer Agol, E.~\& Krolik, J.\ 1999, \apj, 524, 49 

\refer Anders E.~\& Grevesse, N.\ 1989, Geochimica et Cosmochimica Acta 53, 197

\refer Anders E.~\& Ebihara, M.\ 1982, Geochimica et Cosmochimica Acta 46, 2363

\refer Angonin-Willaime, M. -C., Vanderriest, C., Courbin, F., Burud,
I., Magain, P., and Rigaut, F. 1999, Astron. \& Astrophys. 347, 434

\refer Arnaud, K. A. 1996, ASP Conf. Ser. 101:
Astronomical Data Analysis Software and Systems V, ed. G. Jacoby \& J. Barnes (San Francisco: ASP), 17

\refer Blandford, R. D., \& McKee, C. F. 1982, ApJ, 255, 419 

\refer Cash, W.\ 1979, \apj, 228, 939 

\refer Chartas, G., Dai, X., Gallagher, S.~C., Garmire, G.~P., 
Bautz, M.~W., Schechter, P.~L., \& Morgan, N.~D.\ 2001, \apj, 558, 119

\refer Chiba, M, 2001, ApJ, astro-ph/0109499

\refer Corrigan, R.~T., Irwin, M.~J., Arnaud, J., Fahlman, G.~G., 
Fletcher, J.~M., Hewett, P.~C., Hewitt, J.~N., 
Le Fevre, O., McClure, R., Pritchet, C.~J., Schneider, D.~P., 
Turner, E.~L., Webster, R.~L., and Yee, H.~K.~C. 1991, \aj, 102, 34

\refer Eracleous, M., Sambruna, R. M., \& Mushotzky, R. F.  2000, ApJ,
537, 654

\refer Fabian, A.~C., Rees, M.~J., Stella, L., \& White, N.~E.\ 1989, 
\mnras, 238, 729

\refer Falco, E. E., Lehar, J., Shapiro, I. I. 1997, AJ, 113, 540 

\refer Feldman, U.\ 1992, Physica Scripta 46, 202



\refer George, I.~M.~\& Fabian, A.~C.\ 1991, \mnras, 249, 352 

\refer Gould, A.~\& Gaudi, B.~S.\ 1997, \apj, 486, 687 

\refer Grieger, B., Kayser, R., \& Refsdal, S.\ 1988, \aap, 194, 54 

\refer Grieger, B., Kayser, R., \& Schramm, T.\ 1991, \aap, 252, 508 

\refer Halpern, J. P. 1982, PhD Thesis, Harvard University

\refer Hewitt, J. N., Turner, E. L., Lawrence, C. R., Schneider, 
D. P., \& Brody, J. P. 1992, AJ, 104, 968 

\refer Irwin, M.~J., Webster, R.~L., Hewett, P.~C., 
Corrigan, R.~T., \& Jedrzejewski, R.~I.\ 1989, \aj, 98, 1989

\refer Iwasawa, K. \& Taniguchi, Y., 1993, \apj, 413, L15


\refer Katz, C.~A., Moore, C.~B., \& Hewitt, J.~N.\ 1997, \apj, 475, 512

\refer Kayser, R., Refsdal, S., \& Stabell, R.\ 1986, \aap, 166, 36 


\refer Laor, A.\ 1991, \apj, 376, 90 

\refer Lawrence. C, R., Elston, R., Januzzi, B. T., \& Turner,
E. L. 1995, AJ, 110, 2570

\refer Magdziarz \& Zdziarski 1995, MNRAS, 273, 837 

\refer Malhotra, S., Rhoads, J.~E., \& Turner, E.~L.\ 1997, \mnras, 288, 138

\refer Matt, G., Perola, G.~C., \& Piro, L.\ 1991, \aap, 247, 25 

\refer Metcalf, R. B. \& Madau, P., 2001, ApJ, astro-ph/0108224

\refer McLeod, B.~A., Bernstein, G.~M., Rieke, M.~J., \& Weedman, D.~W.\ 1998, \aj, 115, 1377

\refer Mineshige, S.~\& Yonehara, A.\ 1999, \pasj, 51, 497 


\refer Nandra, K., George, I.~M., Mushotzky, R.~F., 
Turner, T.~J., \& Yaqoob, T.\ 1997, \apj, 477, 602

\refer Netzer, H., \& Peterson, B. M. 1997, in Astronomical Time
Series, ed. D. Maoz, A.  Sternberg, \& E. M. Liebowitz (Dordrecht:
Kluwer), 85

\refer Ostensen, R., Refsdal, S., Stabell, R., Teuber, J., 
Emanuelsen, P.~I., Festin, L., Florentin-Nielsen, R., 
Gahm, G., Gullbring, E., Grundahl, F., Hjorth, J., 
Jablonski, M., Jaunsen, A.~O., Kaas, A.~A., Karttunen, H., 
Kotilainen, J., Laurikainen, E., Lindgren, H., 
Maehoenen, P., Nilsson, K., Olofsson, G., Olsen, O., 
Pettersen, B.~R., Piirola, V., Sorensen, A.~N., 
Takalo, L., Thomsen, B., Valtaoja, E., Vestergaard, M., 
and Av Vianborg, T. 1996, \aap, 309, 59

\refer Peterson, B. M. 1993, PASP, 105, 247



\refer Reeves, J.~N.~\& Turner, M.~J.~L.\ 2000, \mnras, 316, 234 


\refer Ros, E., Guirado, J.~C., 
Marcaide, J.~M., P{\' e}rez-Torres, M.~A., Falco, E.~E., Mu{\~ n}oz, J.~A., 
Alberdi, A., \& Lara, L.\ 2000, \aap, 362, 845 

\refer Sambruna, R.~M., Eracleous, M., \& Mushotzky, R.~F.\ 1999, \apj, 526, 60

\refer Schechter, P.~L.~\& Moore, C.~B.\ 1993, \aj, 105, 1

\refer Schneider, P., Ehlers, J., \& Falco, E. E., 1992, Gravitational Lensing (New York: Springer) 

\refer Shalyapin, V.~N.\ 2001, Astronomy Letters, 27, 150


\refer Tonry, J. L., \& Kochanek, C. S., 1999, \apj, 117, 2034 

\refer Trotter, C.~S., Winn, J.~N., \& Hewitt, J.~N.\ 2000, \apj, 535, 671

\refer Tsunemi, H., Mori, K., Miyata, E., Baluta, C., 
Burrows, D.~N., Garmire, G.~P., \& Chartas, G.\ 2001, \apj, 554, 496


\refer Wise, M. W., Davis, J. E., Huenemoerder, Houck, J. C., Dewey, D.
Flanagan, K. A., and Baluta, C. 1997,
{\it The MARX 3.0 User Guide, CXC Internal Document}
available at http://space.mit.edu/ASC/MARX/

\refer Witt, H.~J.\ 1990, \aap, 236, 311

\refer Witt, H. J., Kayser, R. \& Refsdal, S., 1993, A\&A, 268, 501.

\refer Witt, H.~J., Mao, S., \& Schechter, P.~L.\ 1995, \apj, 443, 18

\refer Wo{\' z}niak, P. R., Zdziarski, A. A., Smith, D., Madejski, G. M.,
\& Johnson, W. N. 1998, MNRAS, 299, 449

\refer Wo{\' z}niak, 
P.~R., Alard, C., Udalski, A., Szyma{\' n}ski, M., Kubiak, M., 
Pietrzy{\' n}ski, G., \& Zebru{\' n}, K.\ 2000, \apj, 529, 88 

\refer Wyithe, J.~S.~B., Webster, R.~L., \& Turner, E.~L.\ 2000, \mnras, 318, 762

\refer Wyithe, J.~S.~B. \& Turner, E.~L.\ 2001, \mnras, 320, 21 


\refer Yonehara, A., Mineshige, S., Fukue, J., Umemura, M., \& Turner, E.~L.\ 1999, \aap, 343, 41 

\refer Yonehara, A.\ 2001, \apjl, 548, L127

\refer Young, A.~J.~\& Reynolds, C.~S.\ 2000, \apj, 529, 101

\endrefer

\clearpage
\begin{figure*}[t]
\plotfiddle{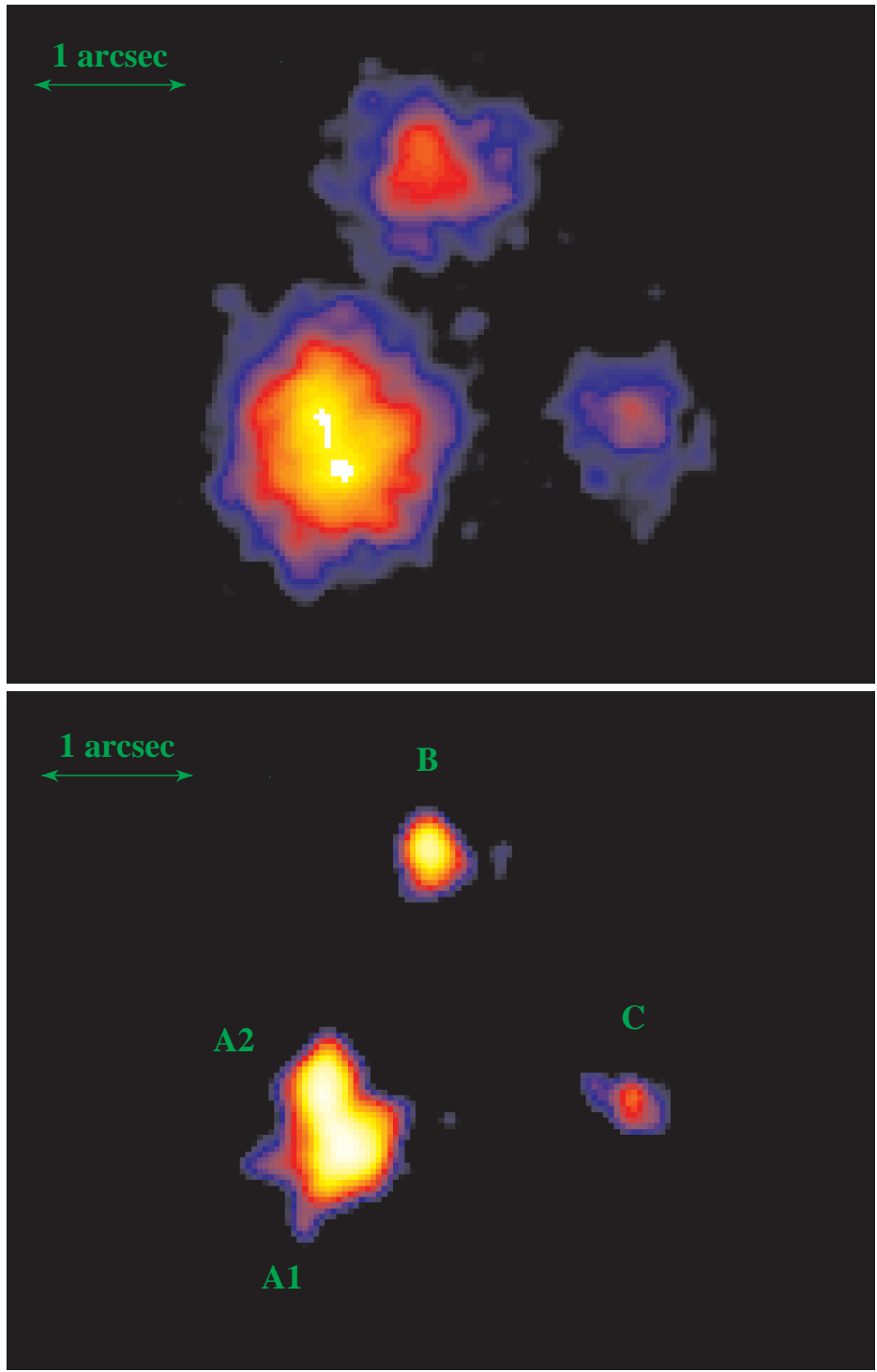}{8.in}{0}{100.}{100.}{-300}{-120}
\protect\caption
{\small (top panel) Combined image of five {\it Chandra} observations of MG~J0414+0534,
(lower panel) Deconvolved image of combined {\it Chandra} observations.
 \label{fig:fig1}}
\end{figure*}

\clearpage
\begin{figure*}[t]
\plotfiddle{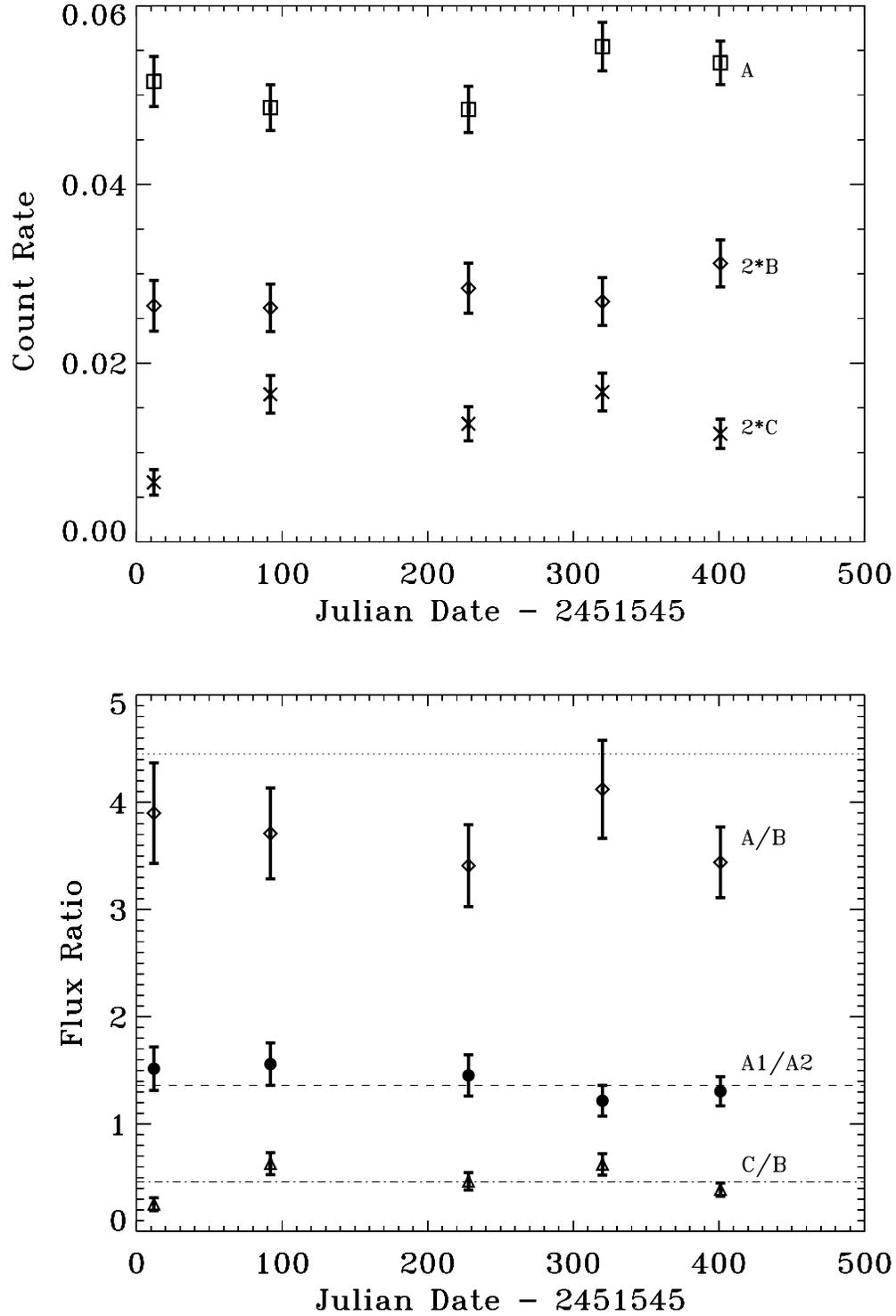}{8.in}{0}{120.}{120.}{-360}{-230}
\protect\caption
{\small (top panel) Light - Curves for Images A, B and C of MG~J0414+0534.
For display purposes the count rates for images B and C have been doubled.
(lower panel) Flux ratios A/B, A1/A2 and C/B. The dotted lines
represent the respective values of the flux ratio in the H-band
obtained from the CfA-Arizona Space Telescope LEns Survey
(CASTLES) web site.
\label{fig:fig2}}
\end{figure*}

\clearpage
\begin{figure*}[t]
\plotfiddle{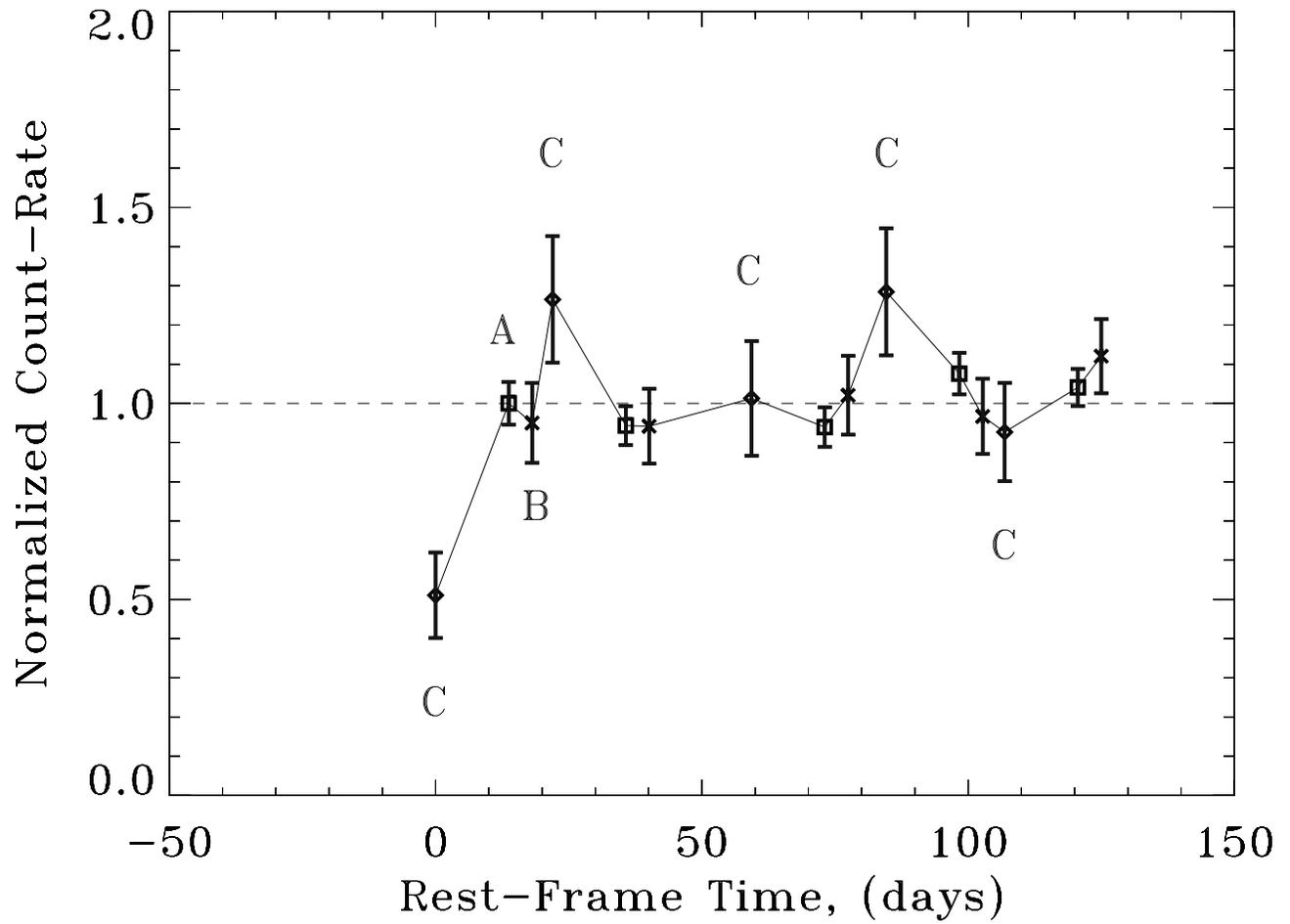}{8.in}{0}{100.}{100.}{-300}{-120}
\protect\caption
{\small Light curve in rest-frame of MG~J0414+0534.
The count-rate of each image is normalized to the mean count-rate of that image
for the five observations.
\label{fig:fig3}}
\end{figure*}

\clearpage
\begin{figure*}[t]
\plotfiddle{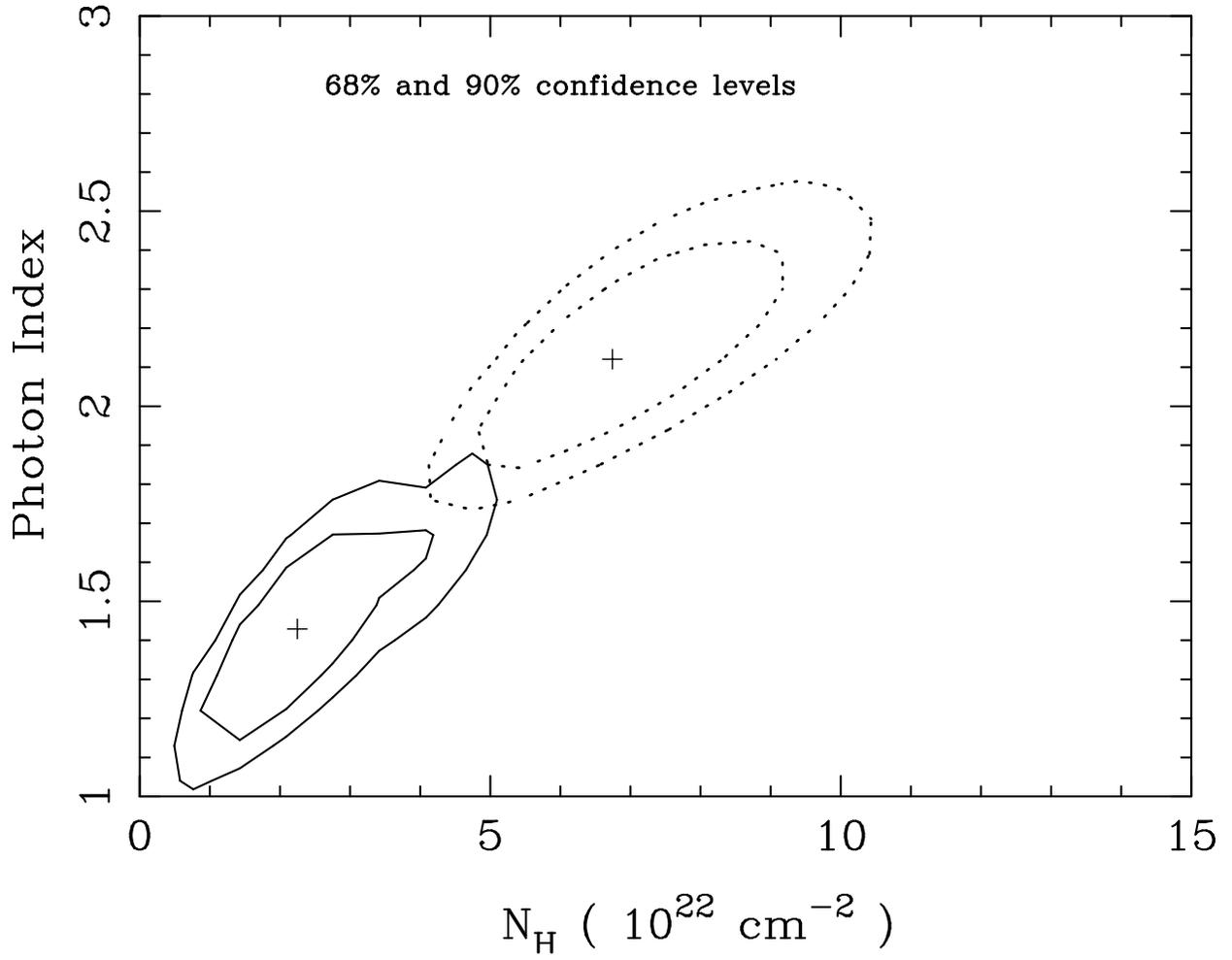}{8.in}{0}{120.}{120.}{-370}{-280}
\protect\caption
{\small 68 \% and 90 \% confidence contours of photon index and intrinsic
absorption in the spectra of image A for the first observation (dotted lines)
and the second observation (solid lines) of MG~J0414+0534.
\label{fig:fig4}}
\end{figure*}

\clearpage
\begin{figure*}[t]
\plotfiddle{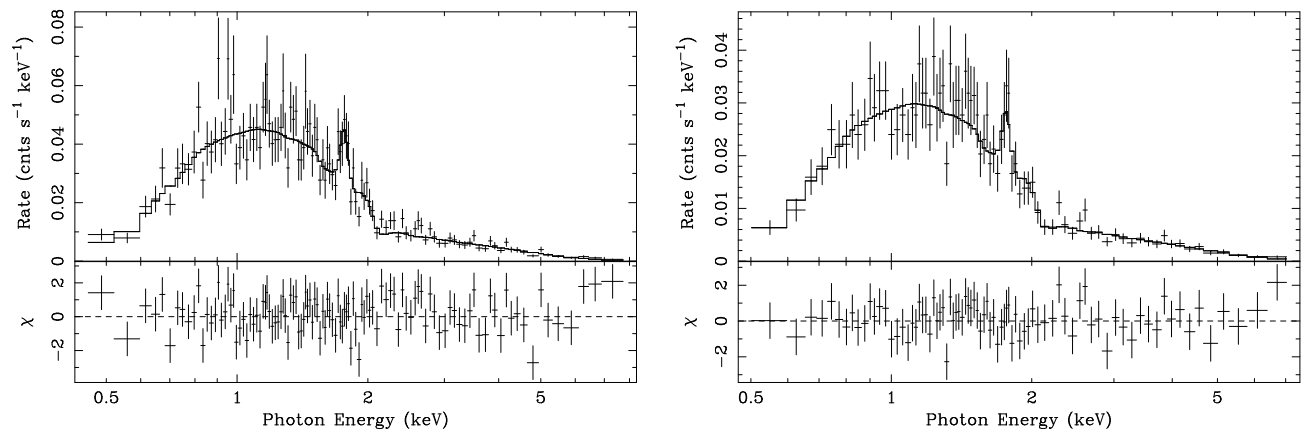}{8.in}{0}{100.}{100.}{-310}{-180}
\protect\caption
{\small (left panel) Combined spectrum of all images of MG~J0414+0534
for the first 5 observations listed in Table 1. (right panel)
Combined spectrum of image A for all observations.
\label{fig:fig5}}
\end{figure*}

\clearpage
\begin{figure*}[t]
\plotfiddle{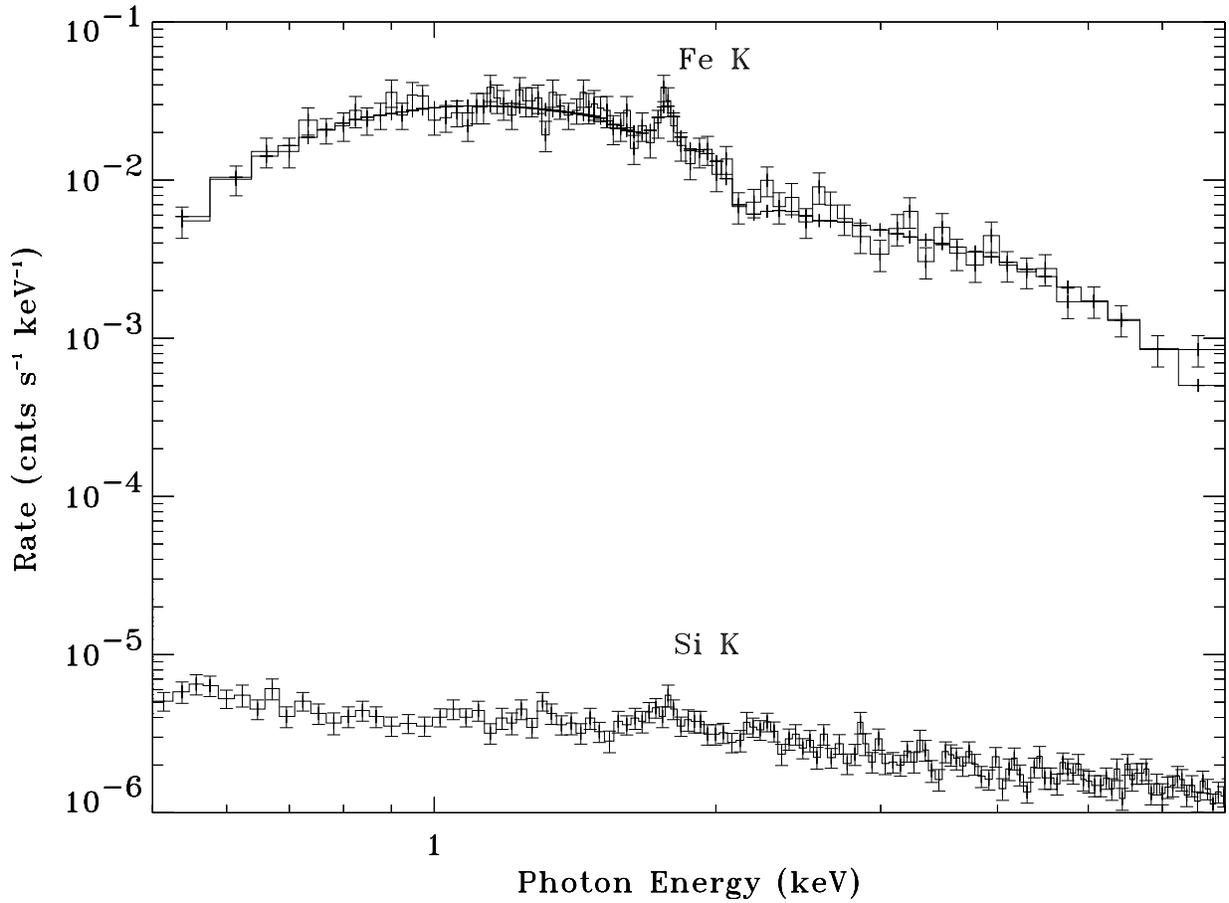}{8.in}{0}{120.}{120.}{-360}{-180}
\protect\caption
{\small Combined spectrum of image A and background for all observations. The
background spectrum is scaled with appropriate normalization to account
for the difference in source and background extraction radii.
The instrumental Si~K line is detected in the background spectrum and is clearly
negligible compared to the detected Fe~K line in the spectrum of image A.
\label{fig:fig6}}
\end{figure*}

\clearpage
\begin{figure*}[t]
\plotfiddle{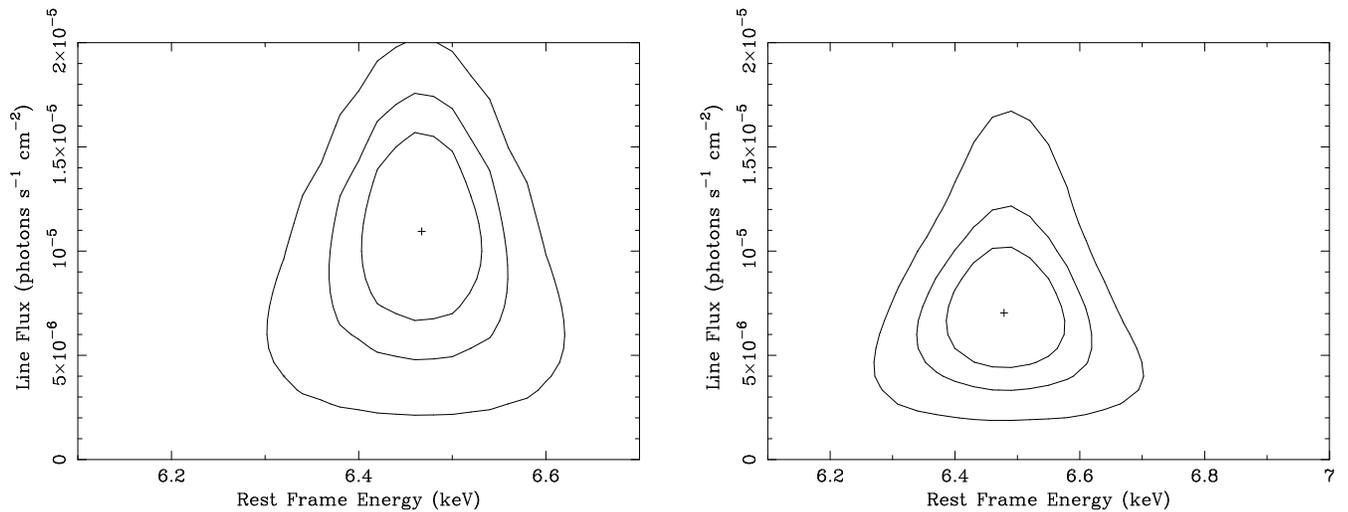}{8.in}{0}{100.}{100.}{-300}{-160}
\protect\caption
{\small 68.3\%, 90\% and 99\% confidence contours of the Fe-K$\alpha$ line
versus the line flux derived for the combined spectrum of all images and observations
of MG~J0414+0534 (left panel) and for the combined spectrum of image B
for observations 4 and 5 (right panel). The mean energy
of the narrow line is consistent with emission from a cold medium.
The broad component that appears in image B can
be satisfactory fit with line emission possibly originating from an ionized accretion disk.
\label{fig:fig7}}
\end{figure*}

\clearpage
\begin{figure*}[t]
\plotfiddle{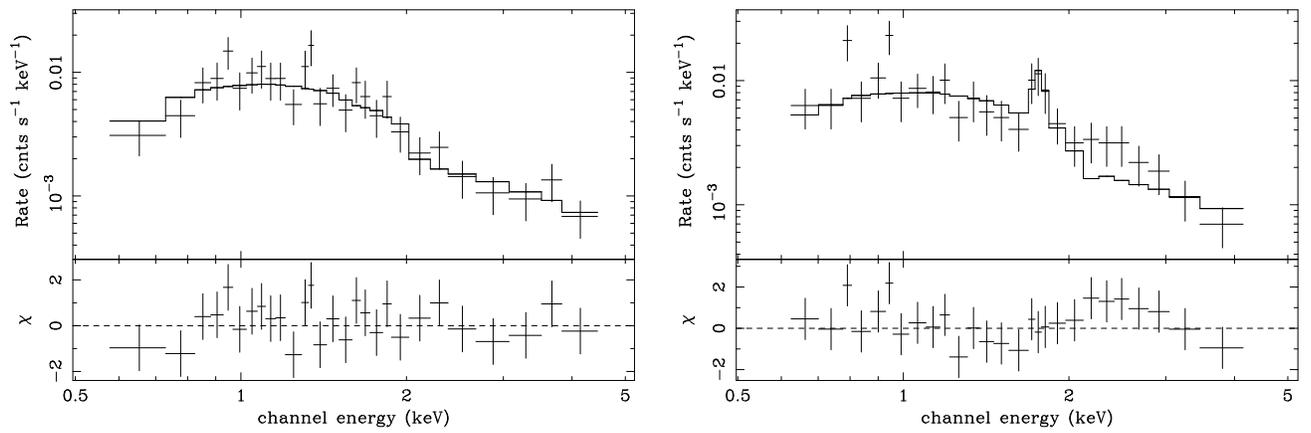}{8.in}{0}{100.}{100.}{-310}{-180}
\protect\caption
{\small (left panel) Combined spectrum of image B of MG~J0414+0534
for the first 3 observations listed in Table 1. (right panel)
Combined spectrum of image B for the last 2 observations listed in Table 1.
\label{fig:fig8}}
\end{figure*}

\clearpage
\begin{figure*}[t]
\plotfiddle{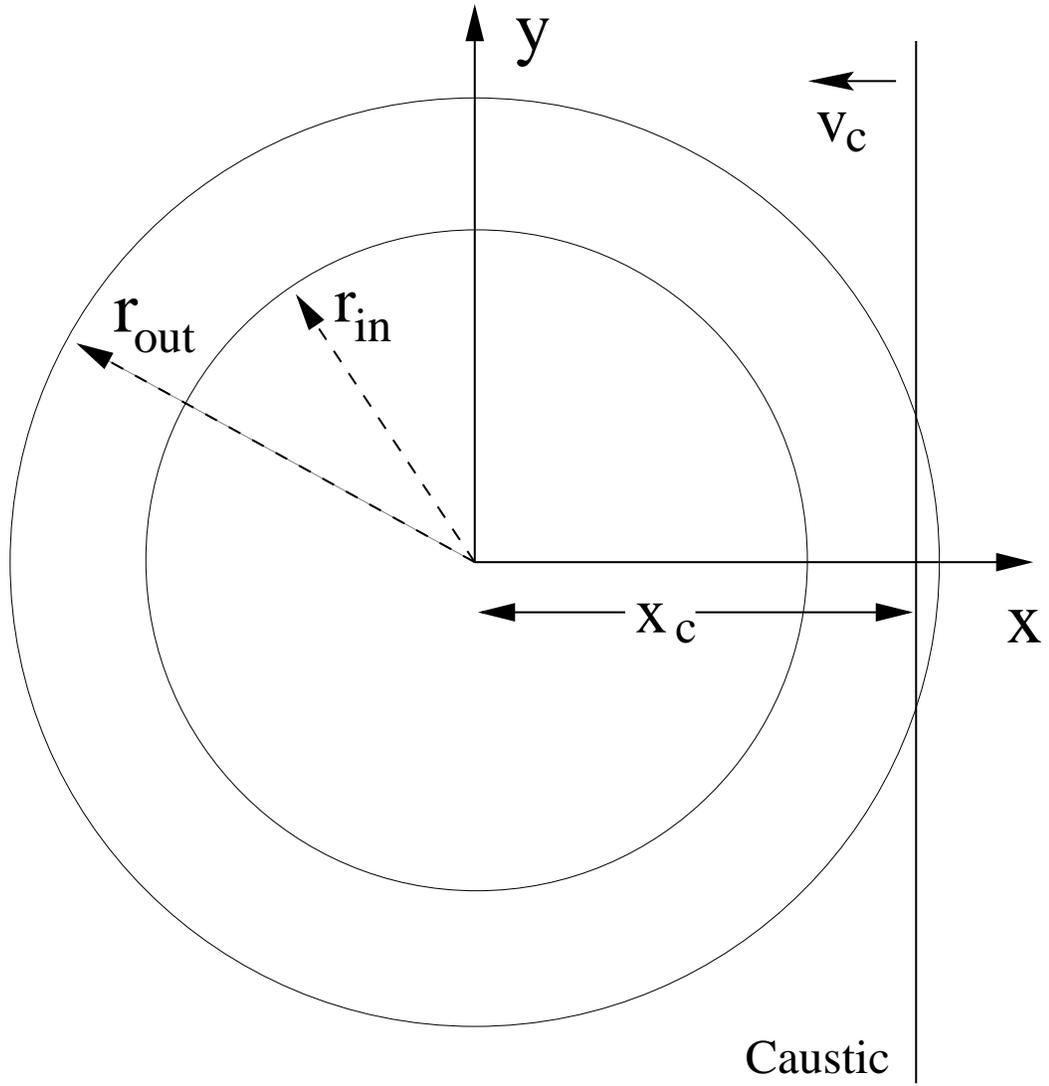}{8.in}{0}{120.}{120.}{-370}{-280}
\protect\caption
{\small Geometry of the accretion disk adopted for our microlensing simulations.
The line parallel to the y axis is the fold custic.
The disk axis points up out of the page.
\label{fig:fig9}}
\end{figure*}

\clearpage
\begin{figure*}[t]
\plotfiddle{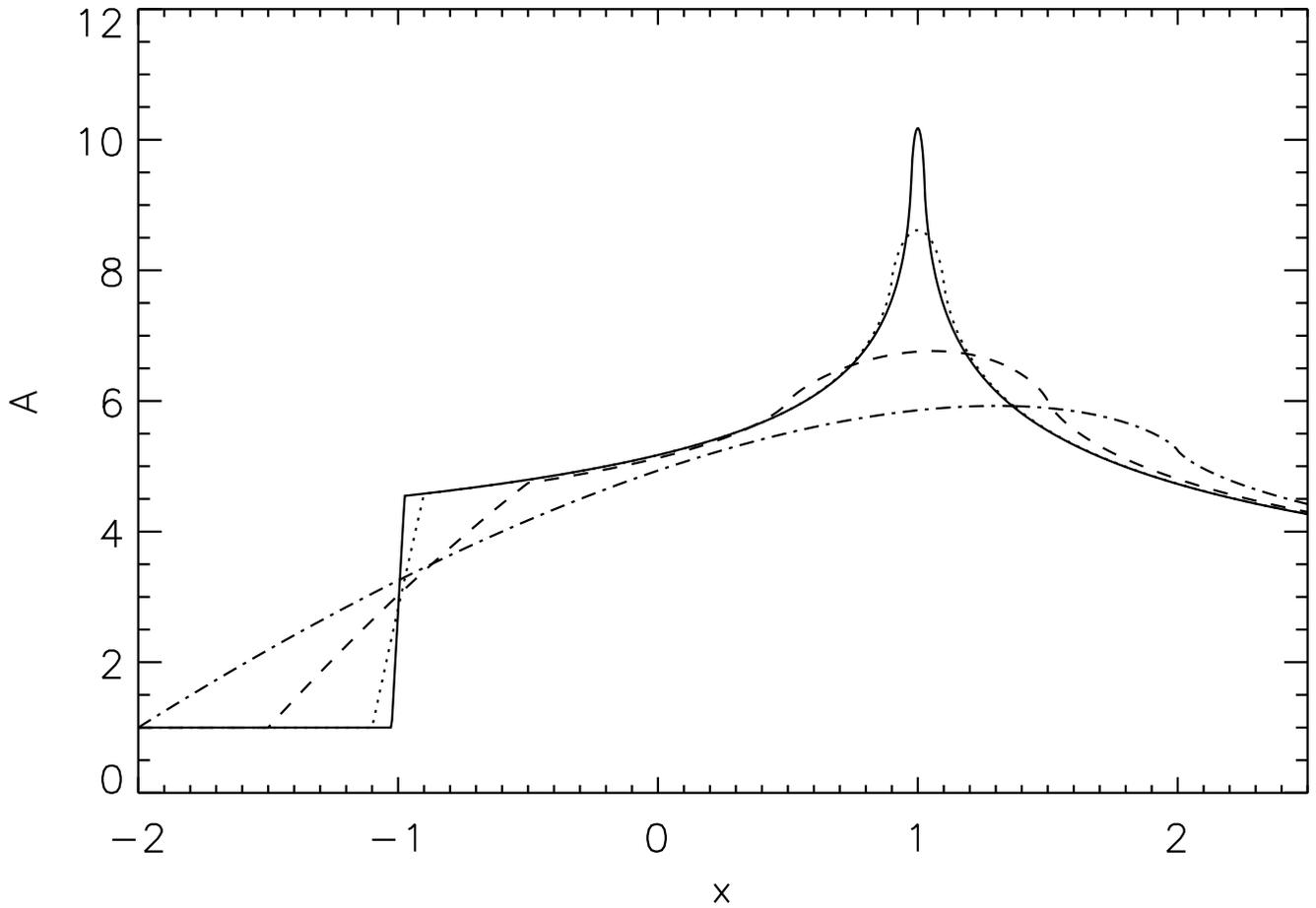}{8.in}{0}{100.}{100.}{-320}{-240}
\protect\caption
{\small The amplification of a ring-shaped
emission region of radius $r$ and thickness $dr$ by a straight fold caustic
with $K=5$ and $A_0=1$ as a function of the distance of the caustic from
the center of the ring. The curves show $dr/r$ = 0.025, 0.1, 0.5, 1 corresponding
to solid, dotted, dashed, and dash-dot lines.
\label{fig:fig10}}
\end{figure*}

\clearpage
\begin{figure*}[t]
\plotfiddle{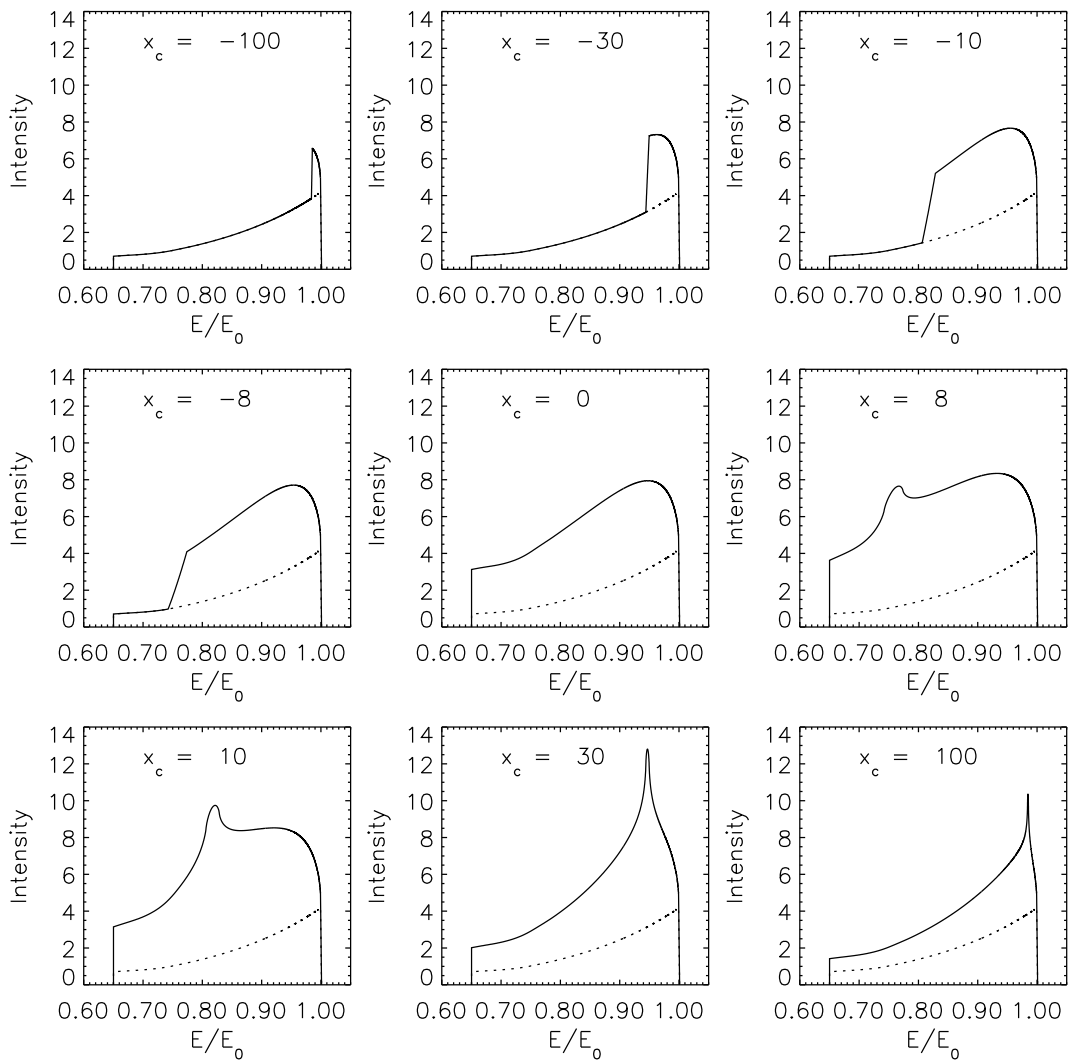}{8.in}{0}{120.}{120.}{-390}{-220}
\protect\caption
{\small
Simulations of the evolution of the Fe K${\alpha}$ line profile as the caustic traverses
a face-on accretion disk around a Schwarzchild black hole. $x_{c}$ is the distance of the caustic from
the center of the accretion disk in units if gravitational radii.
The caustic strength used for the simulations is $K/A_0 = 10r_{g}^{1/2}$.
The assumed velocity broadening of the iron line in the rest frame of the disk is
${\sigma}_{v}/v$ = 0.01, where $v$ is the Keplerian rotation velocity.
\label{fig:fig11}}
\end{figure*}

\end{document}